\pdfoutput=1
\documentclass[12pt,a4paper]{article}
\usepackage{ifthen} 
\newboolean{pdflatex}
\setboolean{pdflatex}{true} 

\newboolean{articletitles}
\setboolean{articletitles}{true} 

\newboolean{uprightparticles}
\setboolean{uprightparticles}{false} 

\newboolean{inbibliography}
\setboolean{inbibliography}{false} 

\def\paperauthors{LHCb collaboration} 
\def\paperasciititle{Measurement of B+, B0 and Lambdab production and nuclear modification in pPb collisions at 8.16 TeV} 
\def\papertitle{Measurement of $\Bp$, $\Bz$ and $\Lb$ production in $p\mkern 1mu\mathrm{Pb}$ collisions at $\sqsnn=8.16\,{\rm TeV}$} 
\def\paperkeywords{$p$Pb collisions, collider, beauty-hadron production, nuclear modifications,  {LHCb}} 
\def\papercopyright{\the\year\ CERN for the benefit of the LHCb collaboration} 
\def\paperlicence{CC-BY-4.0 licence}
\def\paperlicenceurl{https://creativecommons.org/licenses/by/4.0/}

\usepackage[top=1in, bottom=1.25in, left=1in, right=1in]{geometry}

\usepackage{microtype}
\usepackage{lineno}  
\usepackage{xspace} 
\usepackage{caption} 

\usepackage{graphicx}  
\usepackage{color}
\usepackage{colortbl}
\usepackage{booktabs}
\graphicspath{{./figs/}} 
\DeclareGraphicsExtensions{.pdf,.PDF,png,.PNG}

\usepackage{amsmath} 
\usepackage{amssymb}
\usepackage{amsfonts}
\usepackage{upgreek} 

\newcommand*\patchAmsMathEnvironmentForLineno[1]{%
\expandafter\let\csname old#1\expandafter\endcsname\csname #1\endcsname
\expandafter\let\csname oldend#1\expandafter\endcsname\csname
end#1\endcsname
 \renewenvironment{#1}%
   {\linenomath\csname old#1\endcsname}%
   {\csname oldend#1\endcsname\endlinenomath}%
}
\newcommand*\patchBothAmsMathEnvironmentsForLineno[1]{%
  \patchAmsMathEnvironmentForLineno{#1}%
  \patchAmsMathEnvironmentForLineno{#1*}%
}
\AtBeginDocument{%
\patchBothAmsMathEnvironmentsForLineno{equation}%
\patchBothAmsMathEnvironmentsForLineno{align}%
\patchBothAmsMathEnvironmentsForLineno{flalign}%
\patchBothAmsMathEnvironmentsForLineno{alignat}%
\patchBothAmsMathEnvironmentsForLineno{gather}%
\patchBothAmsMathEnvironmentsForLineno{multline}%
\patchBothAmsMathEnvironmentsForLineno{eqnarray}%
}

\usepackage{hyperxmp}

\usepackage[pdftex,
            pdfauthor={\paperauthors},
            pdftitle={\paperasciititle},
            pdfkeywords={\paperkeywords},
            pdfcopyright={Copyright (C) \papercopyright},
            pdflicenseurl={\paperlicenceurl}]{hyperref}

\usepackage[colorinlistoftodos,textsize=scriptsize]{todonotes}

\usepackage[all]{hypcap} 

\usepackage{xspace} 
\usepackage{upgreek}

\def\lhcb   {\mbox{LHCb}\xspace}

\def\MagUp {\mbox{\em Mag\kern -0.05em Up}\xspace}

\ifthenelse{\boolean{uprightparticles}}%
{

 \def\Pmu         {\ensuremath{\upmu}\xspace}

 \def\Ppi         {\ensuremath{\uppi}\xspace}

 \def\Ppsi        {\ensuremath{\uppsi}\xspace}

 \def\PDelta      {\ensuremath{\Delta}\xspace}                 
 \def\PXi         {\ensuremath{\Xi}\xspace}                 
 \def\PLambda     {\ensuremath{\Lambda}\xspace}                 
 \def\PSigma      {\ensuremath{\Sigma}\xspace}                 
 \def\POmega      {\ensuremath{\Omega}\xspace}                 
 \def\PUpsilon    {\ensuremath{\Upsilon}\xspace}

 \def\PB      {\ensuremath{\mathrm{B}}\xspace}                 
                  
 \def\PD      {\ensuremath{\mathrm{D}}\xspace}

 \def\PJ      {\ensuremath{\mathrm{J}}\xspace}                 
 \def\PK      {\ensuremath{\mathrm{K}}\xspace}

 \def\Pb      {\ensuremath{\mathrm{b}}\xspace}                 
 \def\Pc      {\ensuremath{\mathrm{c}}\xspace}

 \def\Pi      {\ensuremath{\mathrm{i}}\xspace}

 \def\Pp      {\ensuremath{\mathrm{p}}\xspace}

 \def\thebaroffset{0.0em}
}
{

 \def\Pmu         {\ensuremath{\mu}\xspace}

 \def\Ppi         {\ensuremath{\pi}\xspace}

 \def\Ppsi        {\ensuremath{\psi}\xspace}                 
                  
 \mathchardef\PDelta="7101
 \mathchardef\PXi="7104
 \mathchardef\PLambda="7103
 \mathchardef\PSigma="7106
 \mathchardef\POmega="710A
 \mathchardef\PUpsilon="7107
                  
 \def\PB      {\ensuremath{B}\xspace}                 
                  
 \def\PD      {\ensuremath{D}\xspace}

 \def\PJ      {\ensuremath{J}\xspace}                 
 \def\PK      {\ensuremath{K}\xspace}

 \def\Pb      {\ensuremath{b}\xspace}                 
 \def\Pc      {\ensuremath{c}\xspace}

 \def\Pi      {\ensuremath{i}\xspace}

 \def\Pp      {\ensuremath{p}\xspace}

 \def\thebaroffset{0.18em}
}
\newcommand{\offsetoverline}[2][\thebaroffset]{\kern #1\overline{\kern -#1 #2}}%

\makeatletter
\ifcase \@ptsize \relax
  \newcommand{\miniscule}{\@setfontsize\miniscule{4}{5}}
\or
  \newcommand{\miniscule}{\@setfontsize\miniscule{5}{6}}
\or
  \newcommand{\miniscule}{\@setfontsize\miniscule{5}{6}}
\fi
\makeatother

\DeclareRobustCommand{\optbar}[1]{\shortstack{{\miniscule (\rule[.5ex]{1.25em}{.18mm})}
  \\ [-.7ex] $#1$}}

\def\mumu       {{\ensuremath{\Pmu^+\Pmu^-}}\xspace}

\def\cquark    {{\ensuremath{\Pc}}\xspace}

\def\bquark    {{\ensuremath{\Pb}}\xspace}

\def\pion   {{\ensuremath{\Ppi}}\xspace}

\def\pip    {{\ensuremath{\pion^+}}\xspace}
\def\pim    {{\ensuremath{\pion^-}}\xspace}

\def\kaon    {{\ensuremath{\PK}}\xspace}

\def\KorKbar {\kern \thebaroffset\optbar{\kern -\thebaroffset \PK}{}\xspace}

\def\Kp      {{\ensuremath{\kaon^+}}\xspace}
\def\Km      {{\ensuremath{\kaon^-}}\xspace}

\def\Kstar   {{\ensuremath{\kaon^*}}\xspace}

\def\Dbar    {{\ensuremath{\offsetoverline{\PD}}}\xspace}
\def\D       {{\ensuremath{\PD}}\xspace}

\def\DorDbar {\kern \thebaroffset\optbar{\kern -\thebaroffset \PD}\xspace}
\def\Dz      {{\ensuremath{\D^0}}\xspace}
\def\Dzb     {{\ensuremath{\Dbar{}^0}}\xspace}

\def\Dm      {{\ensuremath{\D^-}}\xspace}

\def\Dstarzb {{\ensuremath{\Dbar{}^{*0}}}\xspace}

\def\B       {{\ensuremath{\PB}}\xspace}

\def\BorBbar {\kern \thebaroffset\optbar{\kern -\thebaroffset \PB}\xspace}
\def\Bz      {{\ensuremath{\B^0}}\xspace}

\def\Bu      {{\ensuremath{\B^+}}\xspace}

\def\Bp      {{\ensuremath{\Bu}}\xspace}

\def\jpsi     {{\ensuremath{{\PJ\mskip -3mu/\mskip -2mu\Ppsi\mskip 2mu}}}\xspace}

\def\Y#1S{\ensuremath{\PUpsilon{(#1S)}}\xspace}

\def\proton      {{\ensuremath{\Pp}}\xspace}

\def\Lz          {{\ensuremath{\PLambda}}\xspace}

\def\LorLbar     {\kern \thebaroffset\optbar{\kern -\thebaroffset \PLambda}\xspace}

\def\Lc          {{\ensuremath{\Lz^+_\cquark}}\xspace}

\def\Lb           {{\ensuremath{\Lz^0_\bquark}}\xspace}

\def\BF         {{\ensuremath{\mathcal{B}}}\xspace}

\newcommand{\decay}[2]{\mbox{\ensuremath{#1\!\to #2}}\xspace}         

\def\to                 {\ensuremath{\rightarrow}\xspace}

\newcommand{\lqcd}{{\ensuremath{\Lambda_{\mathrm{QCD}}}}\xspace}

\def\AT#1     {\ensuremath{A_{\mathrm{T}}^{#1}}\xspace}           

\def\C#1      {\ensuremath{\mathcal{C}_{#1}}\xspace}                       
\def\Cp#1     {\ensuremath{\mathcal{C}_{#1}^{'}}\xspace}                    
\def\Ceff#1   {\ensuremath{\mathcal{C}_{#1}^{\mathrm{(eff)}}}\xspace}        
\def\Cpeff#1  {\ensuremath{\mathcal{C}_{#1}^{'\mathrm{(eff)}}}\xspace}       
\def\Ope#1    {\ensuremath{\mathcal{O}_{#1}}\xspace}                       
\def\Opep#1   {\ensuremath{\mathcal{O}_{#1}^{'}}\xspace}                    

\newcommand{\nospaceunit}[1]{\ensuremath{\text{#1}}}       
\newcommand{\aunit}[1]{\ensuremath{\text{\,#1}}}       

\newcommand{\tev}{\aunit{Te\kern -0.1em V}\xspace}
\newcommand{\gev}{\aunit{Ge\kern -0.1em V}\xspace}
\newcommand{\mev}{\aunit{Me\kern -0.1em V}\xspace}
\newcommand{\kev}{\aunit{ke\kern -0.1em V}\xspace}
\newcommand{\ev}{\aunit{e\kern -0.1em V}\xspace}
\newcommand{\mevc}{\ensuremath{\aunit{Me\kern -0.1em V\!/}c}\xspace}
\newcommand{\gevc}{\ensuremath{\aunit{Ge\kern -0.1em V\!/}c}\xspace}
\newcommand{\mevcc}{\ensuremath{\aunit{Me\kern -0.1em V\!/}c^2}\xspace}
\newcommand{\gevcc}{\ensuremath{\aunit{Ge\kern -0.1em V\!/}c^2}\xspace}

\def\mum  {\ensuremath{\,\upmu\nospaceunit{m}}\xspace}

\def\mub{\ensuremath{\,\upmu\nospaceunit{b}}\xspace}
\def\nb {\aunit{nb}\xspace}
\def\invnb {\ensuremath{\nb^{-1}}\xspace}

\newcommand{\chisq}{\ensuremath{\chi^2}\xspace}

\newcommand{\chisqip}{\ensuremath{\chi^2_{\text{IP}}}\xspace}

\def\gsim{{~\raise.15em\hbox{$>$}\kern-.85em
          \lower.35em\hbox{$\sim$}~}\xspace}
\def\lsim{{~\raise.15em\hbox{$<$}\kern-.85em
          \lower.35em\hbox{$\sim$}~}\xspace}

\def\sPlot{\mbox{\em sPlot}\xspace}

\def\sqs   {\ensuremath{\protect\sqrt{s}}\xspace}
\def\sqsnn {\ensuremath{\protect\sqrt{s_{\scriptscriptstyle\text{NN}}}}\xspace}
\def\pt         {\ensuremath{p_{\mathrm{T}}}\xspace}

\def\ptot       {\ensuremath{p}\xspace}

\newcommand{\lum} {\ensuremath{\mathcal{L}}\xspace}

\def\evtgen     {\mbox{\textsc{EvtGen}}\xspace}

\def\geant      {\mbox{\textsc{Geant4}}\xspace}

\def\photos     {\mbox{\textsc{Photos}}\xspace}

\xspace

\def\tell1  {TELL1\xspace}
\def\ukl1   {UKL1\xspace}

\newcommand{\ie}{\mbox{\itshape i.e.}\xspace}

\usepackage{cite} 
\usepackage{mciteplus}

\def\pPb {\ensuremath{p\mkern 1mu\mathrm{Pb}}\xspace}
\def\Pbp {\ensuremath{\mathrm{Pb}\mkern 1.5mu p}\xspace}
\def\ystar {\ensuremath{y}\xspace}
\def\decayBpJpsiK {\decay{\Bp}{\jpsi\Kp}}
\def\decayBpDzpi {\decay{\Bp}{\Dzb\pip}}
\def\decayLbLcpi {\decay{\Lb}{\Lc\pim}}
\def\decayBzDppi {\decay{\Bz}{\Dm\pip}}
\def\Hbar {{\ensuremath{\offsetoverline{H}}}\xspace}
\def\Hb {{\ensuremath{H_\bquark}}\xspace}
\def\Hbbar {{\ensuremath{\Hbar{}_\bquark}}\xspace}

\begin{document}

\renewcommand{\thefootnote}{\fnsymbol{footnote}}
\setcounter{footnote}{1}


\begin{titlepage}
\pagenumbering{roman}

\vspace*{-1.5cm}
\centerline{\large EUROPEAN ORGANIZATION FOR NUCLEAR RESEARCH (CERN)}
\vspace*{1.5cm}
\noindent
\begin{tabular*}{\linewidth}{lc@{\extracolsep{\fill}}r@{\extracolsep{0pt}}}
\ifthenelse{\boolean{pdflatex}}
{\vspace*{-1.5cm}\mbox{\!\!\!\includegraphics[width=.14\textwidth]{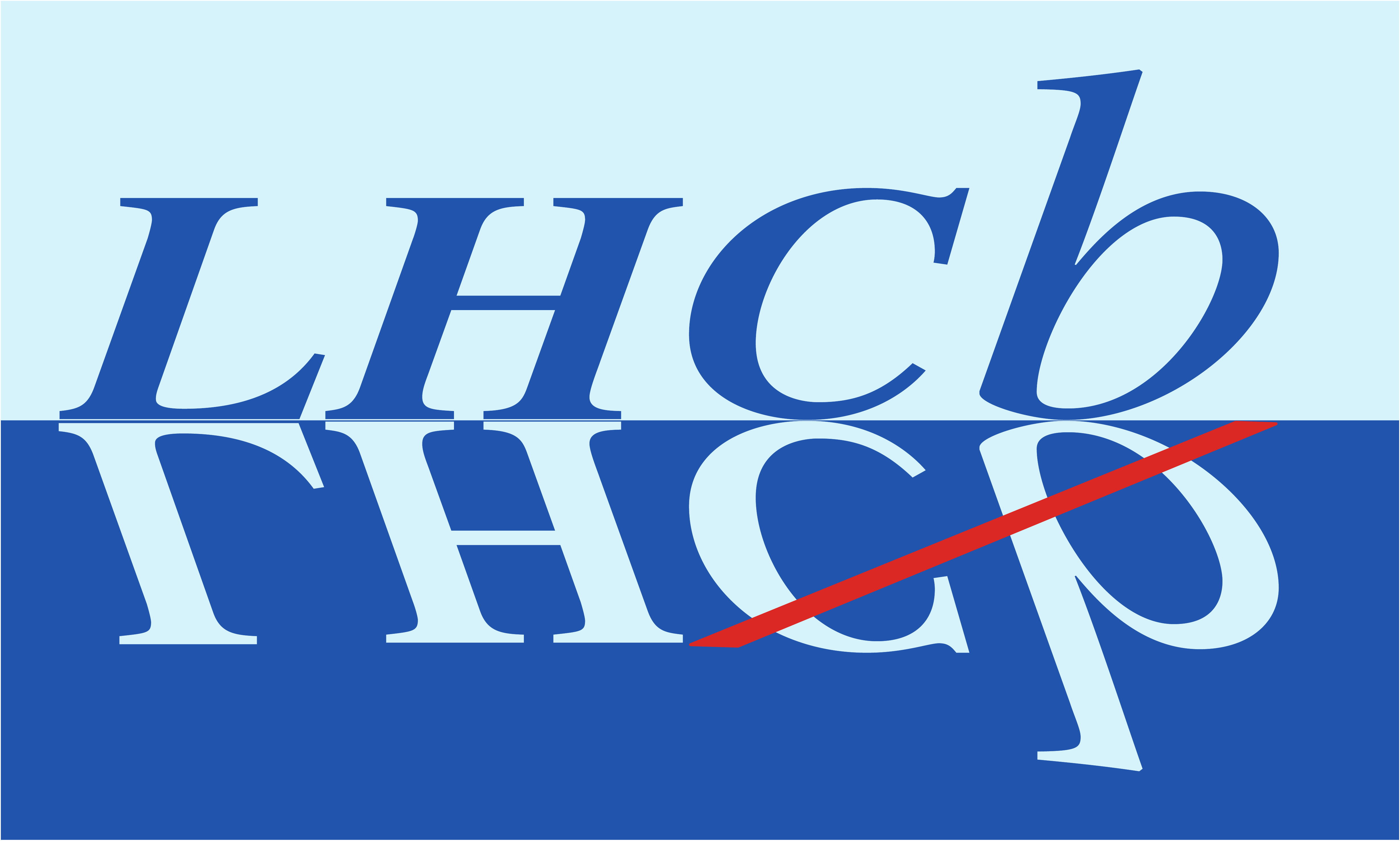}} & &}%
{\vspace*{-1.2cm}\mbox{\!\!\!\includegraphics[width=.12\textwidth]{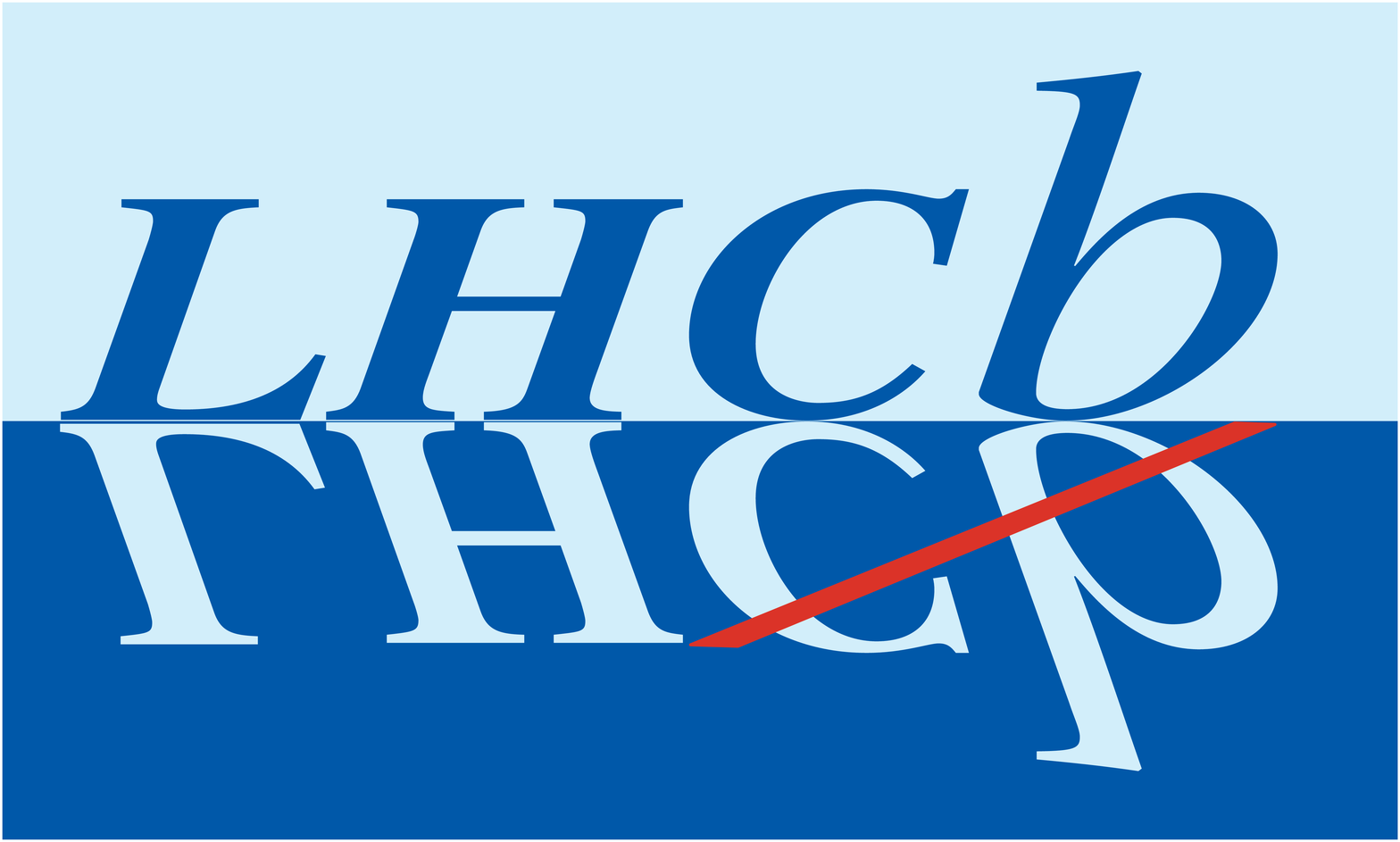}} & &}%
\\
 & & CERN-EP-2019-010 \\  
 & & LHCb-PAPER-2018-048 \\  
 & & 14 February 2019\\ 
 & & \\
\end{tabular*}

\vspace*{2.0cm}

{\normalfont\bfseries\boldmath\huge
\begin{center}
  \papertitle 
\end{center}
}

\vspace*{1.cm}

\begin{center}
\paperauthors\footnote{Authors are listed at the end of this paper.}
\end{center}

\vspace{\fill}

\begin{abstract}
  \noindent
The production of $\Bp$, $\Bz$ and $\Lb$ hadrons is studied in proton-lead collisions
at a centre-of-mass energy per nucleon pair of $\sqsnn=8.16\tev$ recorded with the LHCb detector at the LHC. 
The measurement uses a dataset corresponding to an integrated luminosity of $12.2\pm0.3\invnb$ for the case where the proton beam is projected into the LHCb
detector (corresponding to measuring hadron production at positive rapidity) and $18.6\pm0.5\invnb$ for the lead beam projected
into the LHCb detector (corresponding to measuring hadron production at negative rapidity).  
Double-differential cross-sections are measured and used to determine forward-backward ratios and nuclear modification
factors, which directly probe nuclear effects in the production of beauty hadrons.
The double-differential cross-sections are measured as a function of the beauty-hadron transverse momentum and rapidity in the 
nucleon-nucleon centre-of-mass frame.
Forward-to-backward cross-section ratios and nuclear modification factors indicate a significant nuclear suppression at
    positive rapidity. The ratio of $\Lb$ over $\Bz$ production cross-sections is reported and 
is consistent with the corresponding measurement in $pp$~collisions.  
\end{abstract}

\vspace*{2.0cm}

\begin{center}
    Submitted to Phys.~Rev.~D99 (2019) 052011 
\end{center}

\vspace{\fill}

{\footnotesize 
\centerline{\copyright~\papercopyright. \href{\paperlicenceurl}{\paperlicence}.}}
\vspace*{2mm}

\end{titlepage}


\newpage
\setcounter{page}{2}
\mbox{~}

\cleardoublepage


\renewcommand{\thefootnote}{\arabic{footnote}}
\setcounter{footnote}{0}


\pagestyle{plain} 
\setcounter{page}{1}
\pagenumbering{arabic}

\section{Introduction}
\label{sec:Introduction}

Charm and beauty quarks provide a unique probe of nuclear matter in heavy-ion collisions~\cite{Andronic:2015wma}.
They are produced 
at early times of the collisions and experience the whole evolution of the nuclear medium before hadronization~\cite{Bedjidian:2004gd}. 
Their kinematics and hadronization provide
information on the extent of thermalization effects and on transport coefficients. 
The hard scale provided by the heavy-quark mass is larger than the Quantum Chromodynamics (QCD) scale, \lqcd. 
Therefore, heavy-quark production can be addressed with perturbative QCD down to zero transverse momentum (\pt).  

The characterization of the extended color-deconfined thermodynamic system, the quark-gluon plasma, using heavy-quark
observables in heavy nucleus-nucleus collisions, requires an understanding of background effects.
Therefore, it is mandatory to identify and constrain other QCD effects that may appear in nuclear collisions. Among
these effects, the modification of the parton distribution
functions~\cite{Hirai:2007sx,Eskola:2009uj,deFlorian:2011fp,Kovarik:2015cma,Eskola:2016oht} or, alternatively, 
the breakdown of collinear factorization in the gluon-dense nuclear wave function~\cite{Gelis:2010nm,Fujii:2006ab} are discussed most extensively.
Besides the modification of the nuclear wave function compared to that of free nucleons, coherent 
gluon radiation at small angles may modify final-state heavy-quark kinematic distributions~\cite{Arleo:2012rs}. 
Furthermore, the nuclear effect that is responsible for the change of hadronization patterns as a function of final-state particle
multiplicities in small collision systems ($pp$ and proton-nucleus collisions), first observed for strange-hadron 
production~\cite{ALICE:2017jyt}, is not yet fully understood. Measurements sensitive to hadronization fractions in the
heavy-flavor sector can contribute to a better understanding. Studies of hadronization in heavy nuclear collisions may help to explain the puzzle of  heavy-flavor hadron collective behaviour that was observed recently in $pp$ and proton-lead
collisions~\cite{Strickland:2018exs, Acharya:2017tgv,Sirunyan:2018toe}. These measurements in small collision systems still require a common
reconciliation  with the global theoretical picture of heavy-ion collisions based on fluid dynamics, or might result in
modifications to the fluid-based description. 

Observables related to charm hadrons have been extensively studied at the high-energy frontier of heavy-ion collisions at RHIC and the LHC~\cite{Andronic:2015wma}. 
Recently, the first measurements of \Lc baryon\footnote{The inclusion of 
charge-conjugated state is implicit throughout unless explicitly noted.} production
in proton-lead collisions have been performed at the LHC~\cite{Acharya:2017kfy,LHCb-PAPER-2018-021}.
The measurements of charm-baryon production were the last
important step towards the evaluation of the total charm production cross-section without relying on assumptions about
charm fragmentation functions based on measurements made before the start of the LHC.\footnote{Other charm baryons have
    a negligible contribution to the total charm production.} 
Beauty hadrons are not yet
explored experimentally to the same extent in heavy-ion collisions due to lower production rates.
Theoretically, computations of the production of beauty hadrons are more reliable than charm hadrons since the larger
beauty-quark mass allows for a better separation of energy scales with respect to \lqcd. 
The LHCb collaboration has recently studied the production of $\jpsi$ mesons from beauty-hadron decays (nonprompt
$\jpsi$) in proton-lead
collisions~\cite{LHCb-PAPER-2017-014}. This measurement is sensitive to beauty-quark production 
down to vanishing transverse momentum with good precision. 

This article presents measurements of the production cross-sections of fully reconstructed
$\Bp$, $\Bz$ and $\Lb$ hadrons in proton-lead collisions recorded by the LHCb experiment, as a function of the 
hadron kinematics down to $\pt=2\gevc$, which is lower than the hadron masses. 
The measurement of heavy-quark production at low \pt helps to constrain the gluon wave function in the nucleus in the small Bjorken $x$
region~\cite{LHCb-PAPER-2017-015,Kusina:2017gkz,Aleedaneshvar:2017bgs,Gauld:2016kpd}, where $x$ is the fraction of the nucleon momentum carried by the
interacting gluon.  
In addition, production measurements of fully reconstructed beauty hadrons in heavy-ion collisions can test
whether the hadronization fractions in nuclear collisions are the same as those measured in $pp$ collisions ~\cite{LHCb-PAPER-2012-037,LHCb-PAPER-2014-004,LHCb-PAPER-2011-018,Zhang:2017pmx}.


\section{Detector, data samples and observables}

\label{sec:Detector}

The \lhcb detector~\cite{Alves:2008zz,LHCb-DP-2014-002} is a single-arm forward
spectrometer covering the \mbox{pseudorapidity} range $2<\eta <5$,
designed for the study of particles containing \bquark or \cquark
quarks. The detector includes a high-precision tracking system
consisting of a silicon-strip vertex detector (VELO) surrounding the initial beam
interaction region~\cite{LHCb-DP-2014-001}, a large-area silicon-strip detector located
upstream of a dipole magnet with a bending power of about
$4{\mathrm{\,Tm}}$, and three stations of silicon-strip detectors and straw
drift tubes~\cite{LHCb-DP-2017-001} placed downstream of the magnet.
The tracking system provides a measurement of the momentum, \ptot, of charged particles with
a relative uncertainty that varies from 0.5\% at low momentum to 1.0\% at 200\gevc.
The minimum distance of a track to a primary vertex (PV), the impact parameter (IP), 
is measured with a resolution of $(15+29/\pt)\mum$, where \pt is in\,\gevc.
Different types of charged hadrons are distinguished using information
from two ring-imaging Cherenkov detectors~\cite{LHCb-DP-2012-003}. 
Photons, electrons and hadrons are identified by a calorimeter system consisting of
scintillating-pad and preshower detectors, an electromagnetic
calorimeter and a hadronic calorimeter. Muons are identified by a
system composed of alternating layers of iron and multiwire
proportional chambers~\cite{LHCb-DP-2012-002}.

The online event selection is performed by a trigger~\cite{LHCb-DP-2012-004}, which consists of a hardware stage, based on information from the calorimeter and muon
systems, followed by a two-stage software trigger. The first stage of the software trigger selects displaced high-\pt
tracks or pairs of high-\pt muons, while the second stage searches for $\mu^+\mu^-$ pairs consistent with \jpsi decays
and two-, three- or four-track secondary vertices with a full event reconstruction.
Between the two stages of the software trigger, an alignment and calibration of the detector is performed in 
near real-time \cite{LHCb-PROC-2015-011} and updated constants are made
available for the trigger reconstruction. 
The same alignment and calibration information is propagated 
to the offline reconstruction, ensuring consistent and high-quality 
particle identification (PID) information between the trigger and 
offline software. The identical performance of the online and offline 
reconstruction offers the opportunity to perform physics analyses 
directly using the $\mu^+\mu^-$  pairs reconstructed in the trigger 
\cite{LHCb-DP-2012-004,LHCb-DP-2016-001}, which the present analysis also exploits. 

Simulation is required to model the effects of the detector geometrical acceptance and the efficiency of the selection requirements.
In the simulation, minimum bias proton-lead collisions are generated using the EPOS LHC generator~\cite{Pierog:2013ria}. 
Beauty hadrons ($H_b$) are generated in $pp$ collisions at the same center-of-mass energy
using \textsc{Pythia8}~\cite{Sjostrand:2007gs,Sjostrand:2006za} and are embedded in the minimum bias proton-lead collision events.  
Decays of particles are described by \evtgen~\cite{Lange:2001uf}, in which final-state
radiation is generated using \photos~\cite{Golonka:2005pn}. The
interaction of the generated particles with the detector, and its response,
are implemented using the \geant
toolkit~\cite{Allison:2006ve, *Agostinelli:2002hh} as described in
Ref.~\cite{LHCb-PROC-2011-006}.

The measurement of the production of beauty hadrons in this analysis uses
data recorded in 2016 during the LHC proton-lead run at a centre-of-mass energy per nucleon pair of $\sqsnn=8.16\tev$.
The measurement is performed in bins of beauty-hadron $\pt$ and rapidity, $\ystar$. The rapidity is defined in the nucleon-nucleon 
centre-of-mass frame, using the proton beam direction as the direction of the $z$-axis of the coordinate system. Since
the energy per nucleon in the proton beam is larger than in the lead beam, the nucleon-nucleon centre-of-mass
system has a rapidity in the laboratory frame of $0.465$. 
During the data taking in 2016, the LHC provided collisions with two configurations, inverting
the direction of the proton and lead beams. 
The LHCb forward spectrometer covers the positive (negative) rapidity ranges when the proton (lead) beam direction is 
projected into the detector from the interaction region,
denoted as ``$\pPb$" (``$\Pbp$") configuration.

The dataset corresponds to 
an integrated luminosity of $12.2 \pm 0.3\invnb$ for the \pPb configuration and \mbox{$18.6 \pm 0.5\invnb$} for the \Pbp
configuration, calibrated using dedicated luminosity runs~\cite{LHCb-PAPER-2014-047}. 
The double-differential cross-section of the production of a $H_b$ hadron as a function of $\pt$ and $\ystar$ is computed as
\begin{equation}
\frac{{\rm d}^2 \sigma(H_b)}{{\rm d}\pt \, {\rm d} \ystar } \equiv \frac{N(\Hb)+N(\Hbbar)}{\BF(H_b)\cdot\lum\cdot\epsilon \cdot\Delta \pt\cdot \Delta \ystar} \label{eq:xsec}
\end{equation}
where, for a given interval of $\pt$ and $\ystar$,  $N(\Hb)+N(\Hbbar)$ is the sum of the observed signal yields in a particular decay mode and its 
charge-conjugated decay mode,  $\BF(H_b)$ is the product of the branching fractions for the beauty decay and the subsequent charm decay, $\lum$ is the integrated luminosity, and $\epsilon$ is the total detection efficiency of the final state
particles. 
The measurements are carried out in the kinematic range $2<\pt<20\gevc$ and $1.5<\ystar<3.5$ for the $\pPb$ configuration, and 
in the range $2<\pt<20\gevc$ and $-4.5<\ystar<-2.5$ for the $\Pbp$ configuration. The $\pt$ intervals used to study the
efficiency and signal yield are $2$--$4\gevc$, $4$--$7\gevc$, $7$--$12\gevc$ and $12$--$20\gevc$, 
and the rapidity regions are split into two equal size intervals, $-4.5<y<-3.5$ and $-3.5<y<-2.5$ for the \pPb
configuration, and $1.5<y<2.5$ and $2.5<y<3.5$ for the \Pbp configuration.
The range $\pt<2\gevc$
is not considered due to the small signal yield with the current sample. This restriction is not related to any detector
limitation specific to the collision system, but to the limited integrated luminosity and the small production cross-section.

Nuclear effects are quantified by the nuclear modification factor, $R_{\pPb}$,
\begin{equation}\label{eq:rpa}
R_{\pPb} (\pt,\ystar) \equiv \frac{1}{A_{\rm Pb}} \frac{{\rm d}^2 \sigma_{\pPb}(\pt,\ystar)/{\rm d}\pt{\rm d}\ystar}{{\rm d}^2\sigma_{pp}(\pt,\ystar)/{\rm d}\pt{\rm d}\ystar},
\end{equation}
where 
$A_{\rm Pb}=208$ is the mass number of the lead ion, 
${\rm d}^2 \sigma_{\pPb}(\pt,\ystar)/{\rm d}\pt{\rm d}\ystar$ the $H_b$ production cross-section in proton-lead collisions as
defined in Eq.~\eqref{eq:xsec}, and
${\rm d}^2 \sigma_{pp}(\pt,\ystar)/{\rm d}\pt{\rm d}\ystar$ the $H_b$ reference production cross-section in $pp$ collisions at the 
same nucleon-nucleon centre-of-mass energy. In the absence of nuclear effects, the nuclear modification factor is 
equal to unity. 

To quantify the relative forward-to-backward production rates, the forward-to-backward ratio, $R_{\rm FB}$, is measured, which is  the ratio of cross-sections in the positive and negative
$\ystar$ intervals corresponding to the same absolute value range,
\begin{equation}
R_{\rm FB} (\pt,\ystar) \equiv \frac{{\rm d}^2 \sigma_{\pPb}(\pt,+|\ystar|)/{\rm d}\pt{\rm d}\ystar}{{\rm d}^2 \sigma_{\pPb}(\pt,-|\ystar|)/{\rm d}\pt{\rm d}\ystar}.
\end{equation}


\section{Selections, signal yields and efficiency}
\label{sec:rec_fit_eff_syst}
\subsection{Candidate reconstruction and selection}
\label{sec:rec}
The $\Bp$ cross-section is measured in the $\decayBpJpsiK$ mode, with \mbox{$\jpsi\to\mumu$}, and in the purely hadronic mode
$\decayBpDzpi$, with $\Dzb\to\Kp\pim$. The cross-sections of the $\Bz$ and $\Lb$ hadrons are studied in the
hadronic decays $\decayBzDppi$ with $\Dm\to\Kp\pim\pim$ and $\decayLbLcpi$ with $\Lc\to\proton\Km\pip$.

For the $\decayBpDzpi$, $\decayBzDppi$ and $\decayLbLcpi$ hadronic modes, the candidates are reconstructed
from a sample selected by a hardware trigger requiring a minimum activity in the scintillating-pad detector. 
This hardware trigger selection has an efficiency of 100\% for the signal. 
The intermediate charm-hadron candidates are reconstructed using tracks that are identified as
pion, kaon and proton candidates by the LHCb particle identification system~\cite{LHCb-DP-2014-002}.
The tracks used to form the $\Dzb$ ($\Dm$ and $\Lc$) candidates are required to have $\pt>300\mevc$, and at least one of them has to satisfy $\pt>500\mevc$ ($\pt>1000\mevc$). 
They must also have momentum $p>3\gevc$ ($p>10\gevc$ for protons) and pseudorapidity in the range $2<\eta<5$. 
In addition, they are required to be
separated from any primary vertex by requiring $\chisqip>16$, where $\chisqip$ is the difference between the
$\chisq$ values of a given PV reconstructed with and without the considered track. 
The tracks are required to form a vertex of good quality.
Further requirements are imposed to ensure that
this vertex is consistent with charm-hadron decays by requiring a minimum reconstructed decay time and a reconstructed
mass within an interval centred on the known values of the hadron mass~\cite{PDG2018}: $[1834.8,1894.8]\mevcc$,
$[1844.6, 1894.6]\mevcc$ and
$[2268.5, 2304.5]\mevcc$ for $\Dzb$, $\Dm$ and $\Lc$ candidates, respectively. 
Each mass interval corresponds to six times the experimental resolution on the reconstructed mass.
A charm-hadron candidate, inconsistent with originating from the PVs as ensured by the requirement $\chisqip>4$, is then combined 
with a positively identified pion of the appropriate charge 
to form a beauty hadron. This pion is required to have $\pt>500\mevc$ and to be separated from any PV with the condition $\chisqip>16$.
Reconstructed beauty hadrons with a good-quality vertex and a significant displacement from any PV are selected and are further required to 
point back to a PV by imposing $\chisqip<16$. 
The offline selected beauty-hadron candidates are also required to
match an online vertex, reconstructed from two, three or four tracks, with a large
sum of the transverse momenta of the tracks and a significant displacement from the PVs.

The $\Bp$ candidates studied with the $\decayBpJpsiK$ decay are obtained from a data sample that contains
$\jpsi$ candidates reconstructed by the online software trigger~\cite{LHCb-DP-2016-001}. The muons used to 
reconstruct a  $\jpsi$ meson are
identified by the muon detector and information from all subsystems combined by a neural network.  
The $\jpsi$ candidate must have a well-reconstructed vertex, a mass in the range
$[3056.9,3136.9]\mevcc$, and pass the hardware trigger that selects muons with $\pt>500\mevc$.
The $\jpsi$ candidate with a reconstructed decay vertex significantly separated from all PVs 
is combined with a kaon track to form a $\Bp$ candidate. 
The $\Kp$ candidate must be positively identified and is required to have a 
transverse momentum $\pt>500\mevc$ and to be
separated from all PVs with the requirement $\chisqip>16$. The reconstructed $\Bp$ candidate
is required to have a good-quality vertex, be displaced from the PVs and point back to a PV by requiring $\chisqip<16$.

\subsection{Signal yield determination}
\label{sec:fit}
The signal yields for each decay mode are determined from extended unbinned maximum-likelihood fits to their  mass distributions. 
The fits are used to calculate per-candidate weights with the \sPlot method~\cite{Pivk:2004ty}.  
The weights are then used to determine the signal yields in each \pt and \ystar bin. As a cross-check, fits are also performed in individual $\pt$ and $\ystar$ 
bins, and the results are consistent with those obtained using the \sPlot method. 

The signal mass distribution is described by a Crystal Ball (CB) function~\cite{Skwarnicki:1986xj} for the $\decayBpJpsiK$, $\decayBzDppi$ and $\decayLbLcpi$
decays. 
For the $\decayBpDzpi$ decay an additional Gaussian function, which shares the peak position with the CB function, is necessary to achieve a satisfactory fit quality. 
The tail parameters for the CB function and the fractions of 
the CB and the Gaussian components are fixed to values obtained from fits to mass spectra of simulated signal decays. The mean and
width of the Gaussian core in the CB function, and the width of the separate Gaussian component are free parameters determined from data.
The combinatorial background is described by an exponential function, with parameters allowed to vary in the fits.

The contribution of misidentified background from $\decay{\Bp}{\Dzb\Kp}$, $\decay{\Bz}{\Dm\Kp}$ and 
$\decay{\Lb}{\Lc\Km}$ ($\decay{\Bp}{\jpsi\pip}$) decays, where the $K^\pm$ ($\pip$) meson is reconstructed as a $\pi^\pm$ ($\Kp$) candidate, is described by an empirical
function obtained using simulation. Due to the small branching fraction of the misidentified background
compared to the signal and the suppression from the PID requirement, the contribution relative to the signal mode
in the selected sample is expected to be around or below 5\% depending on the decay mode. 

For the $\decayBpDzpi$ decay, the partially reconstructed backgrounds of $B^{0,+}\to \Dbar{}^{*-,0}\pip$ with $\Dstarzb\to \Dzb\gamma$ or
$\Dbar{}^{*-,0}\to \Dzb\pi^{-,0}$, and $B^{0,+}\to \Dzb\rho^{0,+}$ decays with $\rho^{0,+}\to\pip\pi^{-,0}$, where only the $\Dzb\pip$ in the final states are reconstructed, are modelled with polynomials convolved with
a Gaussian resolution function, following the method described in Ref.~\cite{LHCb-PAPER-2017-021}. 
The partially reconstructed backgrounds of $\decay{B^{-,0}}{\Dm\rho^{0,+}}$ and $\decay{\Lb}{\Lc\rho^{-}}$ ($\decay{B^{0,+}}{\jpsi K^{*0,+}}$) decays,  with
$\rho^{0,\pm}\to\pi^{\pm}\pi^{\mp,0}$ ($K^{*0,+}\to\Kp\pi^{-,0}$), in the $\decayBzDppi$ and $\decayLbLcpi$ ($\decayBpJpsiK$) mass distributions 
are described by a threshold function~\cite{Albrecht:1990am} convolved with a Gaussian function to account for resolution
effects. The resolution function is the same as that of the Gaussian kernel for the signal component.

\begin{figure}[!btp]
\begin{center}
\includegraphics[width=0.49\textwidth]{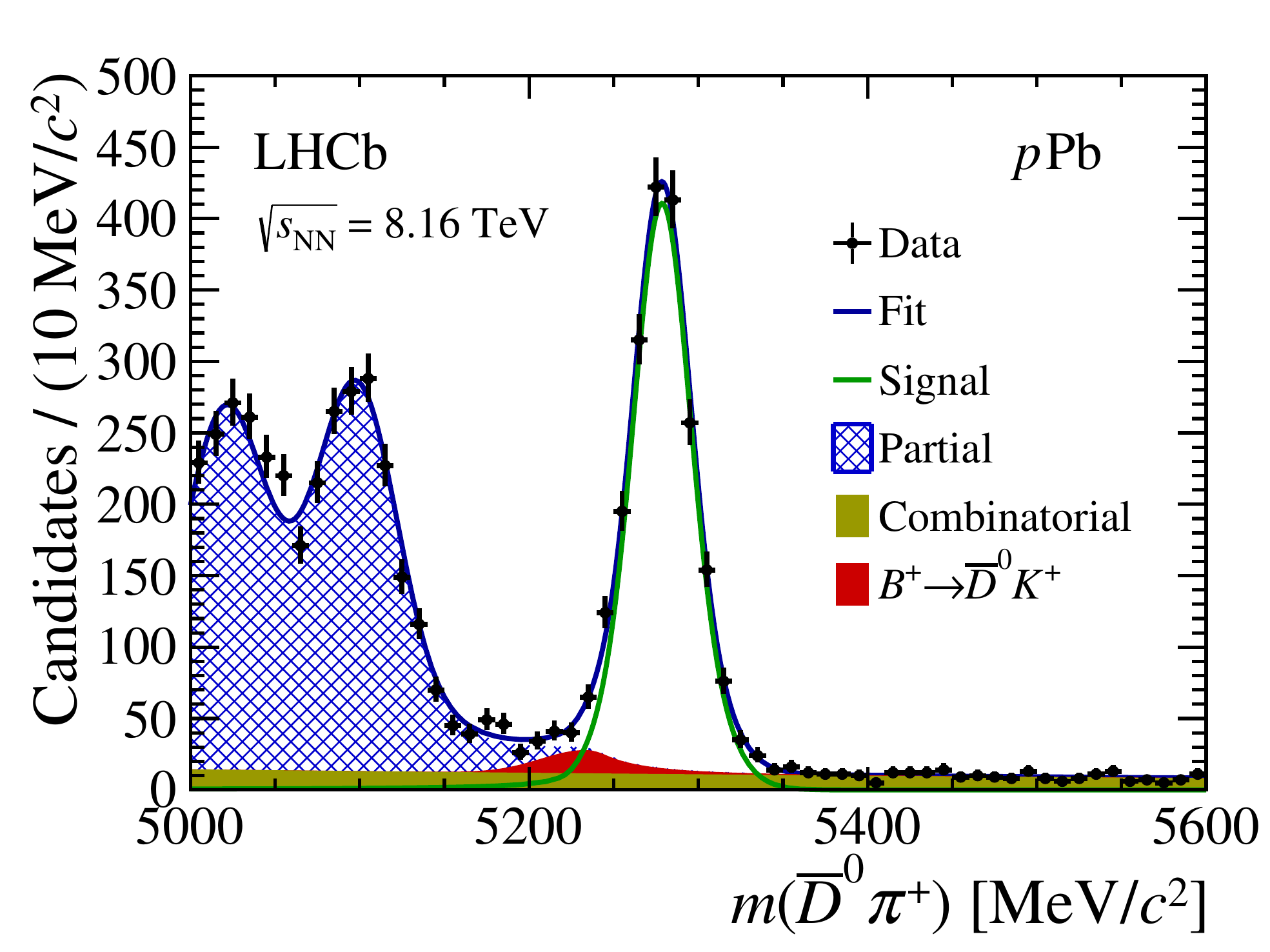}
\includegraphics[width=0.49\textwidth]{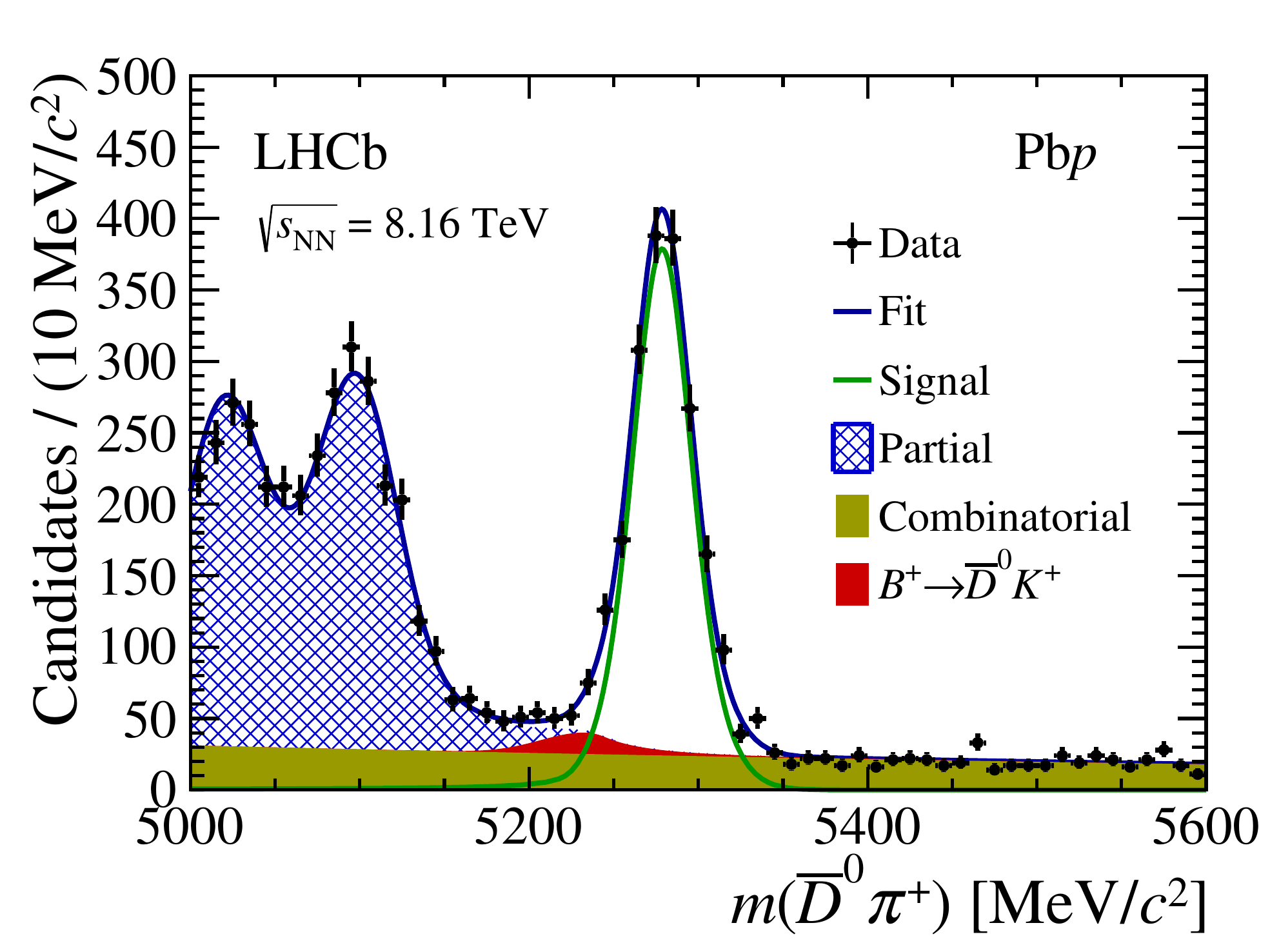}
\end{center}
\caption{
    Invariant mass distribution of $\Bp$ candidates reconstructed in the $\decayBpDzpi$ decay for (left) \pPb and (right) \Pbp collisions, with the fit
    result superimposed.
    The solid blue line, the solid green line, the cross-shaded area, the brown shaded area and the red shaded area
    represent the total fit, the signal component, the partially reconstructed background, the combinatorial background
    and $\decay{\Bp}{\Dzb\Kp}$ decays, respectively.
}
\label{fig:FitBpDzpi}
\end{figure}

\begin{figure}[!btp]
\begin{center}

\includegraphics[width=0.49\textwidth]{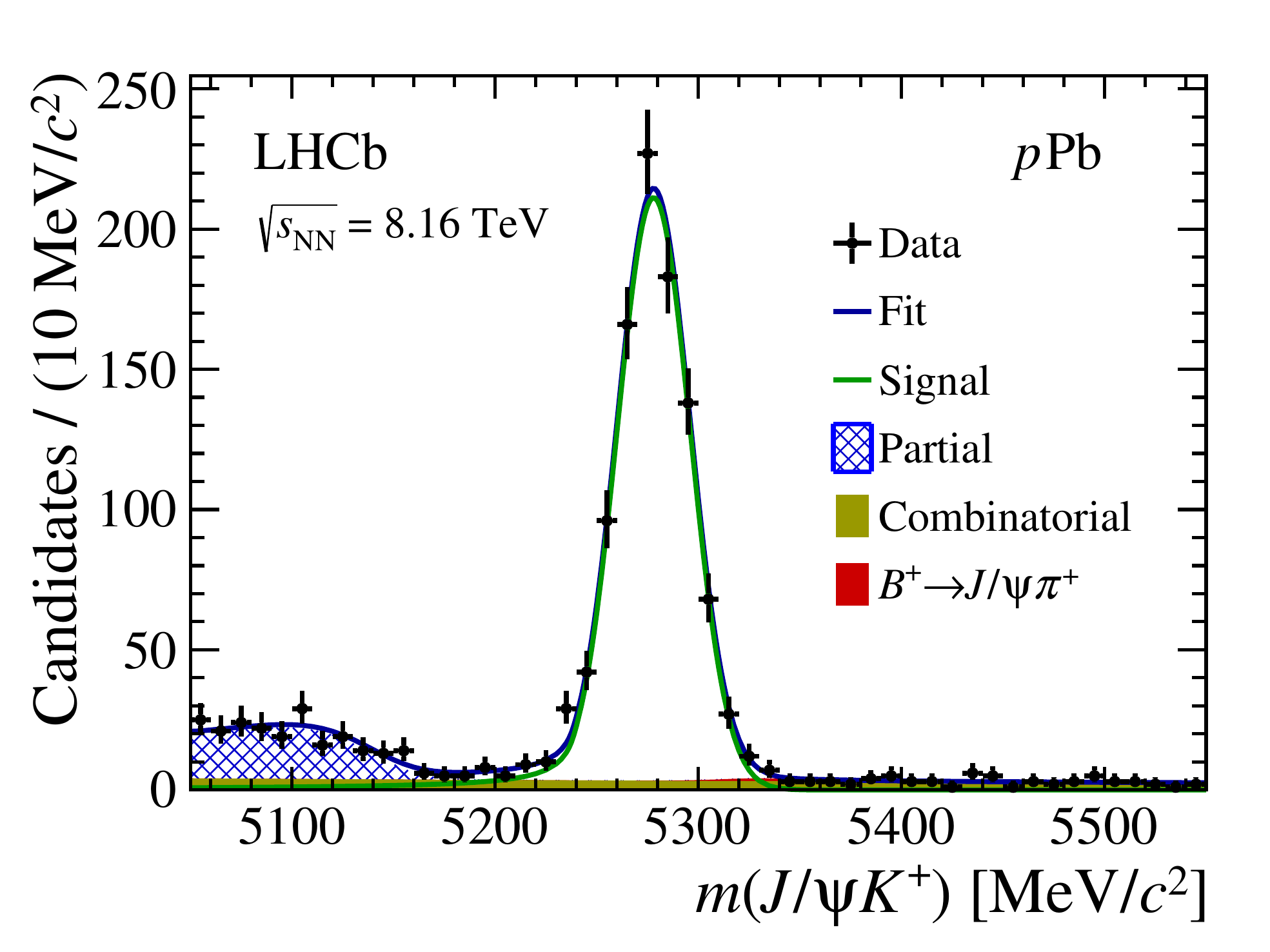}
\includegraphics[width=0.49\textwidth]{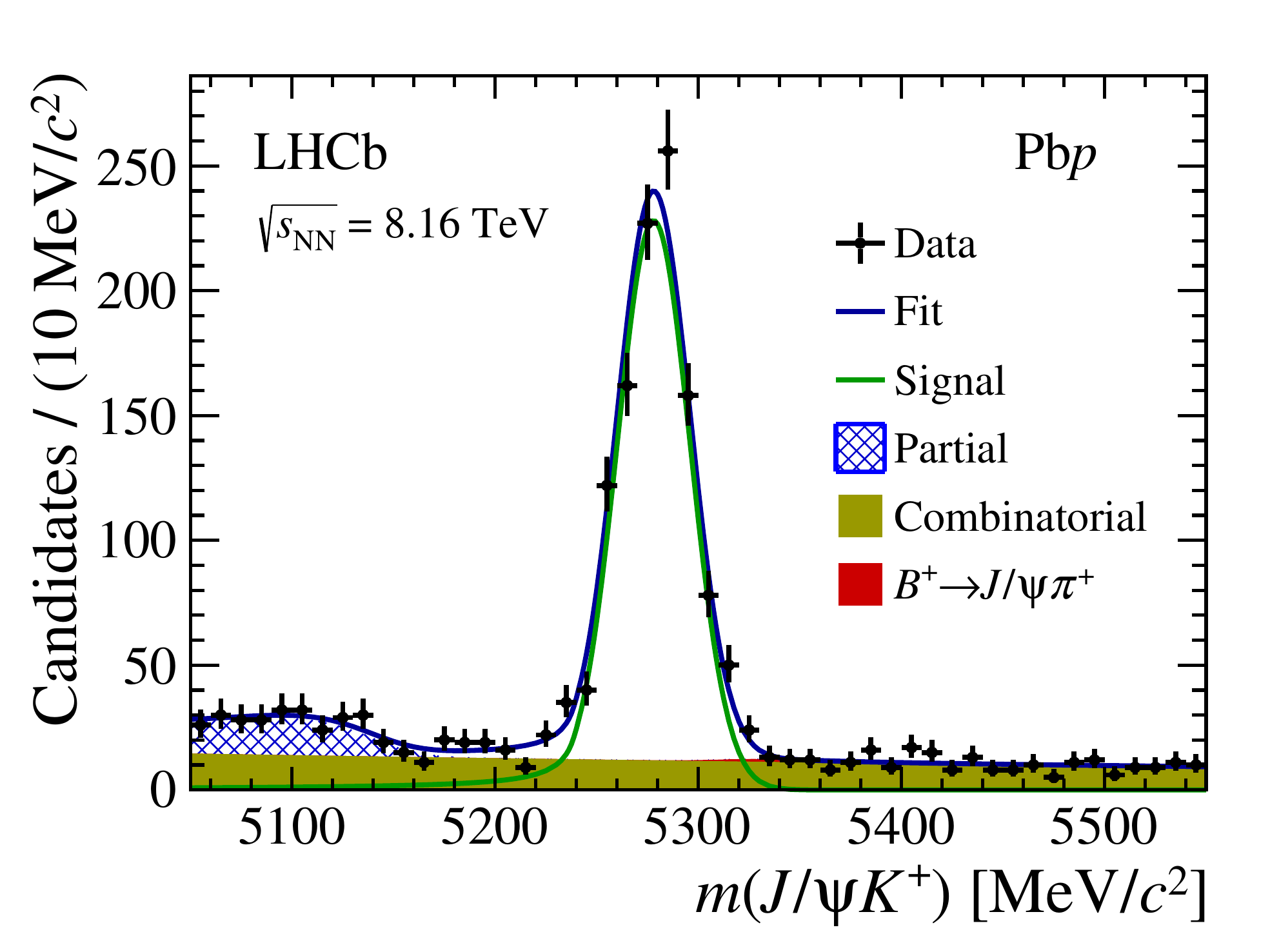}
\end{center}
\caption{
    Invariant mass distribution of $\Bp$ candidates reconstructed in the $\decayBpJpsiK$ decay for (left) \pPb and (right) \Pbp collisions, with the fit
    result superimposed.
    The solid blue line, the solid green line, the cross-shaded area, the brown shaded area and the red shaded area
    represent the total fit, the signal component, the partially reconstructed background, the combinatorial background
    and  $\decay{\Bp}{\jpsi\pip}$ decays, respectively.
}

\label{fig:FitBpJpsiK}
\end{figure}

\begin{figure}[!btp]
\begin{center}
\includegraphics[width=0.49\textwidth]{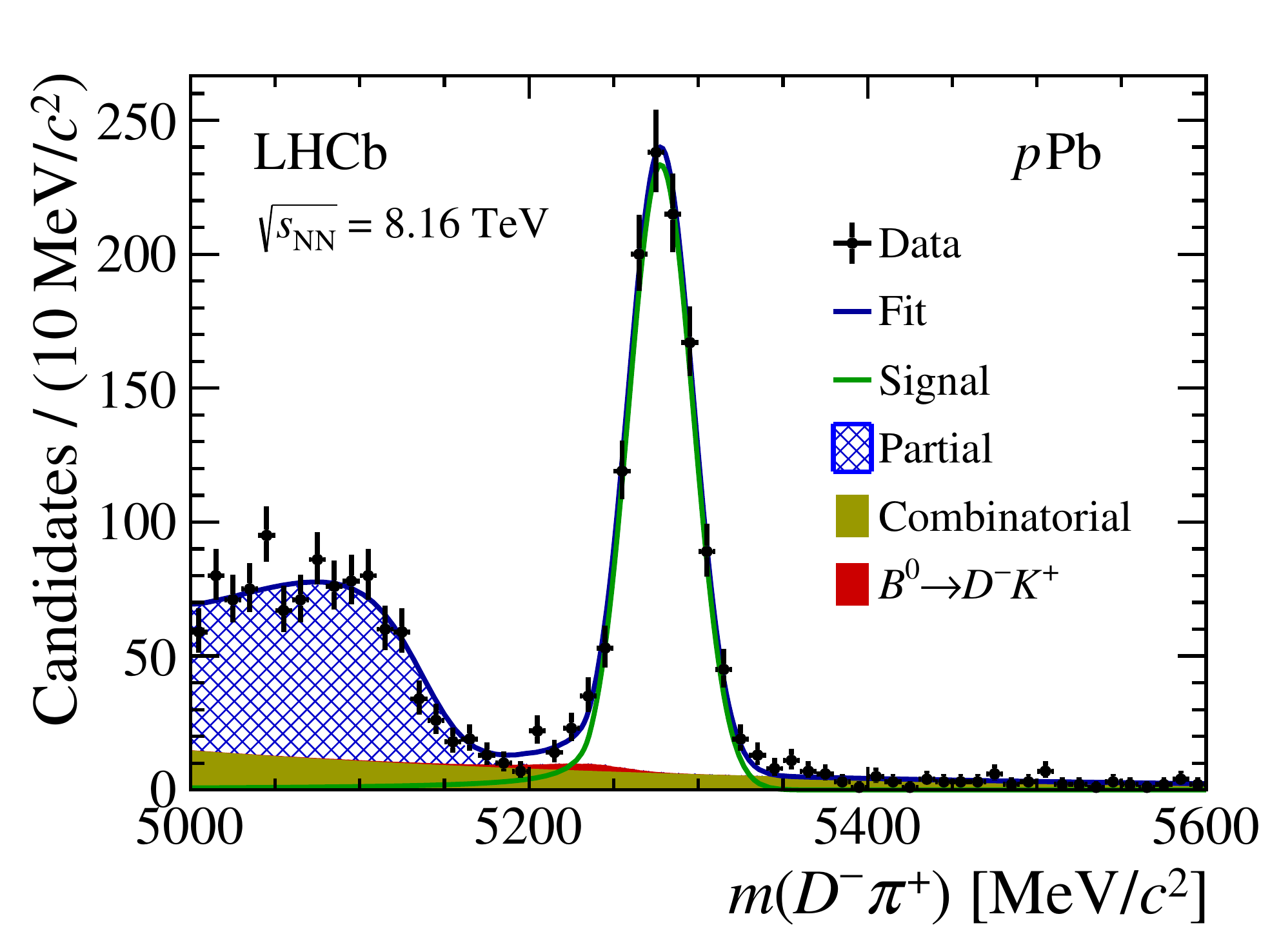}
\includegraphics[width=0.49\textwidth]{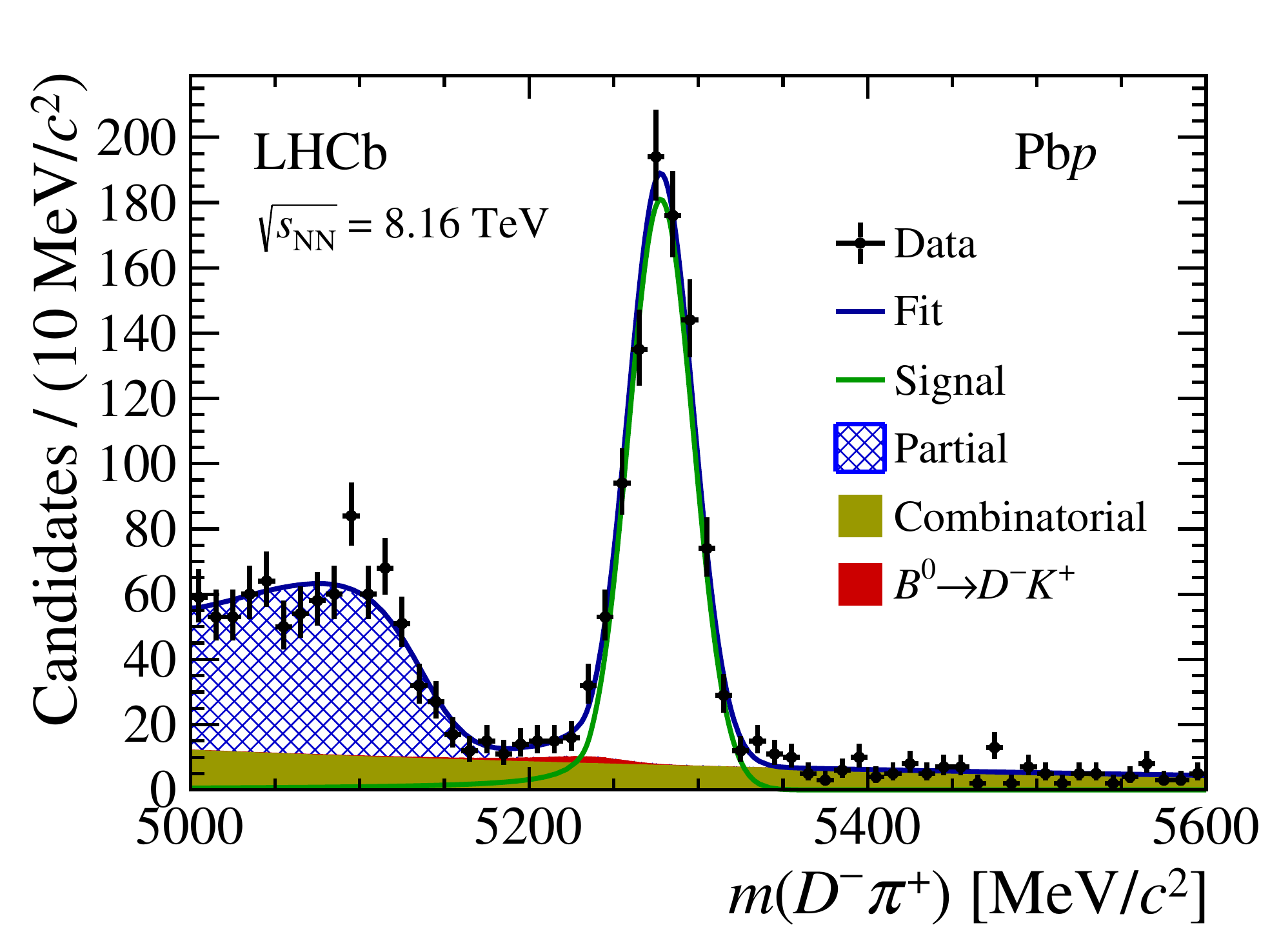}
\end{center}
\caption{
    Invariant mass distribution of $\Bz$ candidates reconstructed in the $\decayBzDppi$ decay for (left) \pPb and (right) \Pbp collisions, with the fit
    result superimposed.
    The solid blue line, the solid green line, the cross-shaded area, the brown shaded area and the red shaded area
    represent the total fit, the signal component, the partially reconstructed background, the combinatorial background
    and $\decay{\Bz}{\Dm\Kp}$ decays, respectively.
}
\label{fig:FitBzDppi}
\end{figure}

\begin{figure}[!btp]
\begin{center}
\includegraphics[width=0.49\textwidth]{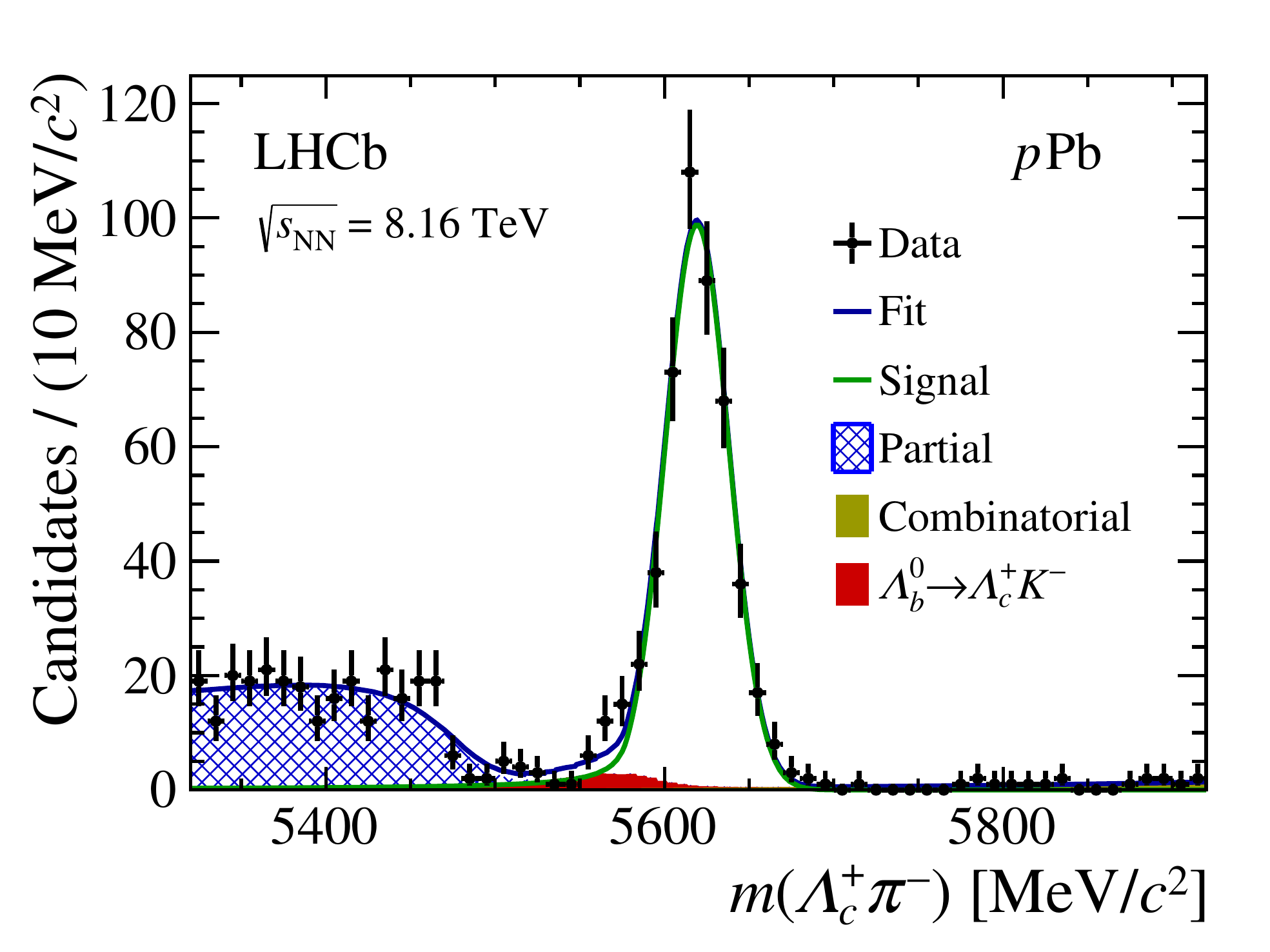}
\includegraphics[width=0.49\textwidth]{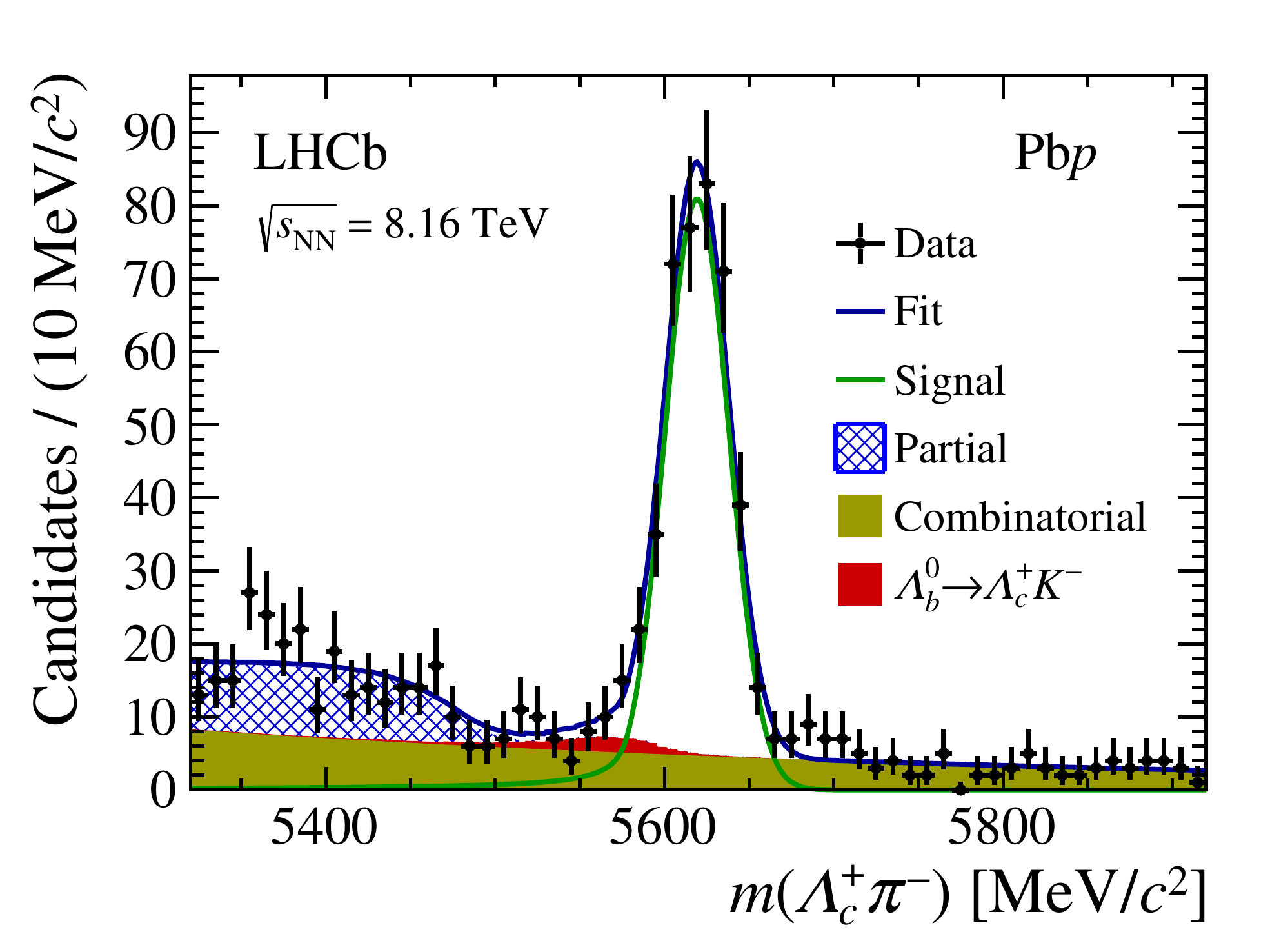}
\end{center}
\caption{
    Invariant mass distribution of $\Lb$ candidates reconstructed in the $\decayLbLcpi$ decay for (left) \pPb and (right) \Pbp collisions, with the fit
    result superimposed.
    The solid blue line, the solid green line, the cross-shaded area, the brown shaded area and the red shaded area
    represent the total fit, the signal component, the partially reconstructed background, the combinatorial background
    and  $\decay{\Lb}{\Lc\Km}$ decays, respectively.
}
\label{fig:FitLbLcpi}
\end{figure}

The shape for each component, except that of the combinatorial background, is constrained to be the same for the fits to
\pPb and \Pbp data. The yields for each contribution in the fit model are free parameters determined from data with the
constraint that the ratio of misidentified background to signal yield is the same in \pPb and \Pbp data.
The signal yields for each decay model considered in this analysis are summarized in Table~\ref{tab:signal_yields} for the kinematic range
$2<\pt<20\gevc$ and $1.5<\ystar<3.5$ ($-4.5<\ystar<-2.5$) in the \pPb (\Pbp) sample. The mass 
distributions and the fit projections are shown in Figs.~\ref{fig:FitBpDzpi} to \ref{fig:FitLbLcpi}
for the decays $\decayBpDzpi$, $\decayBpJpsiK$, $\decayBzDppi$ and $\decayLbLcpi$, respectively. The higher
combinatorial background level in the \Pbp sample compared to the \pPb sample is due to higher charged track
multiplicities seen by the LHCb detector in the \Pbp beam configuration.

\begin{table}[!tbp]
    \caption{Signal yields in the range $2<\pt<20\gevc$ and $1.5<\ystar<3.5$ ($-4.5<\ystar<-2.5$) for
\pPb (\Pbp) collisions. Uncertainties are statistical only.}
\centering
\begin{tabular}{@{}ll@{\:$\pm$\:}ll@{\:$\pm$\:}l@{}}
\toprule
    Decay& \multicolumn{2}{c}{\pPb} & \multicolumn{2}{c}{\Pbp}\\
    \midrule
    $\decayBpDzpi$    &1958 & 54                        &1806 & 55\\
    $\decayBpJpsiK$   &\phantom{0}883 & 32             &\phantom{0}907 & 33\\
    $\decayBzDppi$    &1151 & 38                        &\phantom{0}889 & 34\\ 
    $\decayLbLcpi$    &\phantom{0}484 & 24            &\phantom{0}399 & 23\\
    \bottomrule                                                                                     
\end{tabular}
\label{tab:signal_yields}
\end{table}

\subsection{Efficiency}
\label{sec:eff}
The total efficiency is the product of the geometrical acceptance of the detector and the efficiencies of the reconstruction, the selection, the PID and the trigger requirements.
It is about a few percent in the low-\pt region, and 20\% in the high-\pt region.
These efficiencies, except for the PID, are evaluated using samples of simulated signal decays,
in bins of the beauty-hadron $\pt$ and $\ystar$.
The reconstruction efficiency obtained from simulated signals is corrected using a data-driven method which is detailed in the next paragraph.
The occupancy distribution 
in the minimum bias simulation sample is weighted to reproduce that in data, in order to model correctly the PV reconstruction efficiency. 
For the decays $\decayBpDzpi$, $\decayBpJpsiK$ and $\decayBzDppi$ and subsequent
charm-hadron decays, the angular distributions of the final state particles are well described by \evtgen. For the
$\decayLbLcpi$ decay, the Dalitz-plot distribution of the $\Lc\to\proton\Km\pip$ decay in  simulation  is described by a
mixture of uniform phase space and resonant contributions of $\Delta(1232)^{++}\to \proton\pip$ and $\Kstar (892)^0\to \Km\pip$. 
The $\Lc$ Dalitz-plot distribution in the simulation is corrected to match that in the background subtracted data.

The track reconstruction efficiency from simulation is corrected using a tag-and-probe
approach. For this method, $\jpsi$ candidates in data are formed combining a fully reconstructed ``tag'' track with a
``probe'' track reconstructed using a subset of the tracking detectors~\cite{LHCb-DP-2013-002,LHCb-DP-2014-002}. 
The single-track reconstruction efficiency is obtained as the fraction of the probe tracks that are matched to fully reconstructed
tracks, in bins of the track momentum and pseudorapidity. The ratio of the tag-and-probe efficiency between proton-lead data and simulation is used to
correct the simulation efficiencies. The correction factors are determined for the \pPb and \Pbp  samples
separately.  

The PID efficiency for each track is determined with a tag-and-probe method~\cite{LHCb-PUB-2016-021,LHCb-DP-2018-001}
using calibration samples of proton-lead data. 
The track PID efficiency depends on the detector occupancy. Since  the occupancy distribution is found to be consistent
between the calibration samples and the beauty-signal events, the efficiency is parametrized as a function of track momentum and
pseudorapidity. The pion and kaon PID efficiencies are calibrated using
$\Dz\to\Km\pip$ decays, where the $\Dz$ flavor is tagged by the charge of the pion in $D^{*+}\to\Dz\pi^{+}$ decays, the proton PID efficiency is studied using $\Lz\to \proton\pim$ decays and the PID efficiency for
muons is obtained using $\jpsi\to\mumu$ decays. 
For each beauty candidate, the product of the single-track PID efficiencies, measured as a function of the track
momenta and pseudorapidity, gives the combined PID efficiency for all the tracks in the final state. The efficiency is then averaged over
all beauty-hadron candidates for each bin of $\pt$ and $\ystar$. 


\section{Systematic uncertainties}
\label{sec:systematic}

The various sources of systematic uncertainties, and their quadratic sum, on the cross-sections  for $\Bp$, $\Bz$ and $\Lb$ hadrons are summarized in
Tables~\ref{tab:sys_summary_Fwd} and~\ref{tab:sys_summary_Bwd} for the $\pPb$ and $\Pbp$ data samples, respectively. 
The ranges in the tables correspond to the minimum and maximum values over the \pt and \ystar bins of the
measurement.
The cross-section of the $\Bp$ hadron is measured in the two decay modes, $\decayBpJpsiK$ and $\decayBpDzpi$, 
which give consistent results within statistical uncertainties.

\begin{table}[!tbp]
\caption{
    Summary of systematic uncertainties (in \%) for the measured cross-sections for different decay modes in \pPb. The
    ranges correspond to the minimum and maximum values over the \pt and \ystar bins of the measurement.
   }
\centering
\begin{tabular}{@{}lllll@{}}
    \toprule
    Source                     &$\decayBpJpsiK$& $\decayBpDzpi$ & \decayBzDppi &  \decayLbLcpi\\
    \midrule
    Luminosity                 & 2.6     & 2.6     & 2.6     & 2.6      \\
    Trigger                    & 1.0     & 1.0     & 1.0     & 1.0      \\
    Signal yield               & 2.0     & 2.0     & 2.0     & 2.0      \\
    Selection                  & 1.0     & 1.0     & 3.0     & 2.0      \\
    Hadron tracking            & 1.5     & 4.5     & 6.0     & 6.0      \\
    Tracking efficiency method & 2.4     & 2.4     & 3.2     & 3.2      \\
    Tracking sample size       & $2.0$--$4.3$ & $2.4$--$4.9$ & $3.4$--$9.5$ & $3.3$--$8.0$  \\
    Branching fraction         & 3.1     & 3.2     & 6.0     & 9.6      \\
    PID binning                & $0.0$--$0.7$ & $0.0$--$0.6$ & $0.0$--$0.9$ & $0.1$--$1.4$  \\
    PID sample size            & $1.4$--$2.7$ & $0.2$--$0.6$ & $0.2$--$0.7$ & $0.2$--$0.4$  \\
    Kinematics                 & $0.1$--$4.1$ & $0.5$--$5.4$ & $0.1$--$7.0$ & $0.2$--$9.4$  \\
    Dalitz structure           & --      & --      & --      & $0.8$--$3.1$  \\
    Simulation sample size  & $0.7$--$2.2$ & $0.8$--$2.4$ & $1.4$--$3.7$ & $0.9$--$4.1$  \\
    \midrule
    Total                   &$6.3$--$8.1$  & $7.5$--$10.3$ & $10.9$--$14.5$ & $13.1$--$18.3$\\ 
    \bottomrule
\end{tabular}
\label{tab:sys_summary_Fwd}
\end{table}
\begin{table}[!tbp]
\caption{
    Summary of systematic uncertainties  (in \%) for the measured cross-sections for different decay modes in \Pbp. The
    ranges correspond to the minimum and maximum values over the \pt and \ystar bins of the measurement.
   }
\centering
\begin{tabular}{@{}lllll@{}}
    \toprule
    Source                     &$\decayBpJpsiK$& $\decayBpDzpi$ & \decayBzDppi &  \decayLbLcpi\\
    \midrule
    Luminosity                 & 2.5       & 2.5       & 2.5       & 2.5      \\
    Trigger                    & 1.0       & 1.0       & 1.0       & 1.0      \\
    Signal yield               & 2.0       & 2.0       & 2.0       & 2.0      \\
    Selection                  & 1.0       & 1.0       & 3.0       & 2.0      \\
    Hadron tracking            & 1.5       & 4.5       & 6.0       & 6.0      \\
    Tracking efficiency method & 2.4       & 2.4       & 3.2       & 3.2      \\
    Tracking sample size       & $4.6$--$11.1$  & $5.4$--$10.5$  & $7.8$--$17.8$  & $7.7$--$14.7$ \\
    Branching fraction         & 3.1       & 3.2       & 6.0       & 9.6      \\
    PID binning                & $0.0$--$1.0$   & $0.1$--$0.7$   & $0.0$--$0.6$   & $0.1$--$1.4$  \\
    PID sample size            & $0.7$--$2.1$   & $0.1$--$0.4$   & $0.2$--$0.5$   & $0.1$--$0.2$  \\
    Kinematics                 & $0.7$--$3.9$   & $0.1$--$2.5$   & $0.5$--$1.9$   & $0.3$--$6.9$  \\
    Dalitz structure           & --        & --        & --        & $0.8$--$3.1$  \\
    Simulation sample size     & $0.8$--$2.6$   & $1.1$--$2.7$   & $1.9$--$3.8$   & $1.9$--$3.9$  \\
     \midrule
     Total                   &$7.4$--$12.7$  & $9.0$--$13.1$ & $13.0$--$20.9$ & $15.1$--$21.3$\\ 
    \bottomrule
\end{tabular}
\label{tab:sys_summary_Bwd}
\end{table}

The uncertainty on the $b$-hadron signal yields is studied by using alternative fit models or 
different fitting ranges for the  mass distributions.  The nominal CB function for the signal mass 
distribution is replaced by a combination of a Gaussian function plus a CB function or vice-versa for the $\decayBpDzpi$ decay,
giving a relative change of 2\% on the signal yields for all the decay modes.
A second-order polynomial is employed to replace the exponential function for the combinatorial background, which
results in a difference of 1\% for the signal yields at the maximum.
The effect of partially reconstructed background is studied by fitting the mass distribution in a smaller region where its
contribution is reduced or absent. The signal yields change by at most 1\% for all the decay channels.
The effect of the misidentified background is studied by fixing its branching fraction relative to that of the
signal~\cite{PDG2018}, corrected by the PID selection efficiency. The change in signal yields amounts to 0.1\%.
The maximum value among all these effects, 2\%, is quoted as the systematic uncertainty,  and is considered as a
global uncertainty for all decay modes and all \pt and \ystar bins.

The corrections to the track reconstruction efficiency are limited in precision by the size of the calibration data
sample, which results in a systematic uncertainty dominating in most of the analysis bins.  This effect 
is studied by generating sets of correction factors according to Gaussian distributions centered on their nominal
values and with widths equal to the statistical uncertainties. The standard deviation of the variations of the corrected efficiency in simulation is assigned
as uncertainty,  labelled as ``Tracking sample size'' in the summary tables. It ranges from 2.0\% to 9.5\% for  \pPb
and from 4.6\% to 17.8\% for  \Pbp, depending on the decay modes and the beauty-hadron \pt and \ystar bins. The larger
uncertainty for the \Pbp sample, where the LHCb detector accepts particles produced in the lead beam direction, is due
to higher background that makes the signal yield determination in the calibration data sample more difficult.
The tag-and-probe method used to calculate the tracking efficiency has an uncertainty
estimated to be 0.8\% per track~\cite{LHCb-DP-2013-002}, giving a total value of 2.4\% (3.2\%) for a three- (four-)track
decay mode. Since the tracking efficiency is measured using muons, an additional uncertainty
of 1.5\% per track is introduced for hadrons, to account for the possible imperfect modeling of the amount of 
interactions with the detector material. Labelled as ``Hadron tracking'' in the summary tables, the result is equal to $1.5\%$ for 
$\decayBpJpsiK$ and to 4.5\% (6\%) for three- (four-)track hadronic decays. The uncertainties related to the track reconstruction 
efficiency method and to the hadron-detector interactions are fully correlated among different hadron species and between
the \pPb, \Pbp and $pp$ datasets.

Several sources of systematic uncertainties are associated with the PID efficiencies.
The contribution due to the limited  size of the data calibration samples is determined by varying the single-track PID efficiencies 
within their uncertainties for all momentum and pseudorapidity bins simultaneously, and calculating the resulting spread of the PID 
efficiencies on the $b$-hadron signal decays. Since large samples are available for the kaon, pion, and proton calibration, the resulting systematic uncertainties
are found to be small and in the range of $0.2\%$--$0.7\%$ ($0.1\%$--$0.5\%$) for the $\decayBpDzpi$ decay,
$\decayBzDppi$ and $\decayLbLcpi$ decays in \pPb (\Pbp) collisions. 
They are labelled as ``PID sample size'' in the summary tables. 
For $\decayBpJpsiK$ decays, the smaller size of the muon calibration samples
results in a systematic uncertainty between 1.4\% and 2.7\% for the \pPb data and between 0.7\% and 2.1\% for the \Pbp
data. 
For each bin of track momentum and pseudorapidity, the possible difference in track kinematics between the PID sample
and the  $b$-hadron sample is counted as a second source of systematic uncertainty. The effect is studied by varying the default binning
scheme using finer bins, and determining the changes of the PID efficiencies on the $b$-hadron signal decays. 
The result is labelled as ``PID binning'' in the summary tables and is found to be at most 1.4\%.
The systematic uncertainty related to a possible difference of detector occupancy between the PID samples and the 
$b$-hadron samples is studied by weighting the occupancy in the PID samples to match that of the signal beauty sample, and the resulting change of the efficiency is found to be
negligible.

The imperfect modeling of $b$-hadron kinematic distributions and decay properties in the simulation 
introduces systematic uncertainties on the reconstruction and selection efficiencies.
The two-body invariant mass distributions of the $\Lc$ decay products, or Dalitz-plot distribution, for the $\decayLbLcpi$ mode in simulation is weighted to match data, 
and the uncertainty on the Dalitz-plot distribution is counted as a source of systematic uncertainty. Its magnitude is studied 
by pseudoexperiments. For each pseudoexperiment, a sample is constructed by randomly sampling $\Lb$ candidates from data  allowing for repetition, and this sample is used to correct  the Dalitz-plot distribution in the simulation.
The root-mean-square value of the efficiencies corrected with multiple pseudoexperiments is quoted as the systematic uncertainty. It  is found to
be in the range $0.8\%$--$3.1\%$ for the different $\Lb$ \pt and \ystar bins and is labelled as ``Dalitz structure'' in
the summary tables.

The distributions of variables used to select candidates show good agreement
between data and simulation. The effect of the residual differences is quantified by weighting the reconstructed 
$b$-hadron decay-time distribution in simulation to match that in data,
and studying the corresponding variation of the selection efficiency. The result, labelled as
``Selection'' in the  summary tables, amounts to 1\% for the two $\Bp$ decay modes, and to 3\% and 2\% for the $\Bz$ and
$\Lb$ decay modes.

Simulation and data also show reasonable agreement in the beauty-hadron \pt and \ystar distribution, even if a modest
discrepancy in the \pt distribution is observed, especially for the $\Lb$ baryon.
Due to the limited data sample size it is not possible to accurately determine the $b$-hadron $\pt$ and $\ystar$  distributions from data
directly. However, as the cross-section is measured differentially in bins of \pt and \ystar, the small discrepancy on these kinematic distributions has a reduced impact. 
A systematic uncertainty is evaluated as the change in the reconstruction efficiency after reweighting the \pt and
$\ystar$ distributions in simulation to match data using a finer binning scheme.
The result, labelled as ``Kinematics'' in the summary tables, ranges 
from a fraction of a percent to a few percent depending on the decay modes and the beauty-hadron \pt and \ystar bins.

The muon trigger efficiency is validated using a large sample of
$\jpsi\to\mumu$ decays obtained with an unbiased trigger selection~\cite{LHCb-DP-2012-004}. 
The result is compared with the trigger efficiency estimated in simulation, showing a difference of at most 1\%,
which is quoted as the systematic uncertainty due to the trigger selection for the $\decayBpJpsiK$ decay.
Thanks to the loose requirement applied by the online event selection, the overall trigger
efficiency for the purely hadronic decay modes is found to be above 99\% for the offline selected candidates.  A systematic uncertainty of 1\% is
assigned.

The finite sizes of the simulated $b$-hadron signal samples introduce uncertainties on the efficiency,
which are propagated to the cross-section. Labelled as ``Simulation sample size'', these uncertainties range from subpercent to a few percent depending on the decay modes and the  $\pt$ and $\ystar$ bins.
The uncertainties due to the integrated luminosity of the \pPb and \Pbp datasets are of 2.6\% and 2.5\%, respectively.
The uncertainties on the branching fractions of the $b$-hadron decays and of the intermediate charm-hadron decays are 
also sources of systematic uncertainty, and are evaluated using the uncertainties on the measured values~\cite{PDG2018}.

The dominant systematic effect is the uncertainty on the track reconstruction efficiency which, however, largely cancels in the cross-section ratios.
For the $\decayLbLcpi$ decay, the branching fraction is also a large source of systematic uncertainty, but cancels for
the nuclear modification factor measurements. The systematic uncertainties   are
considered to be fully correlated among all kinematic bins for a particular decay mode, except that labelled as ``Simulation sample size'' which is uncorrelated.


\section{Results}

\subsection{Cross-sections}
\begin{figure}[!tpb]
\centering

  \includegraphics[width=0.49\textwidth]{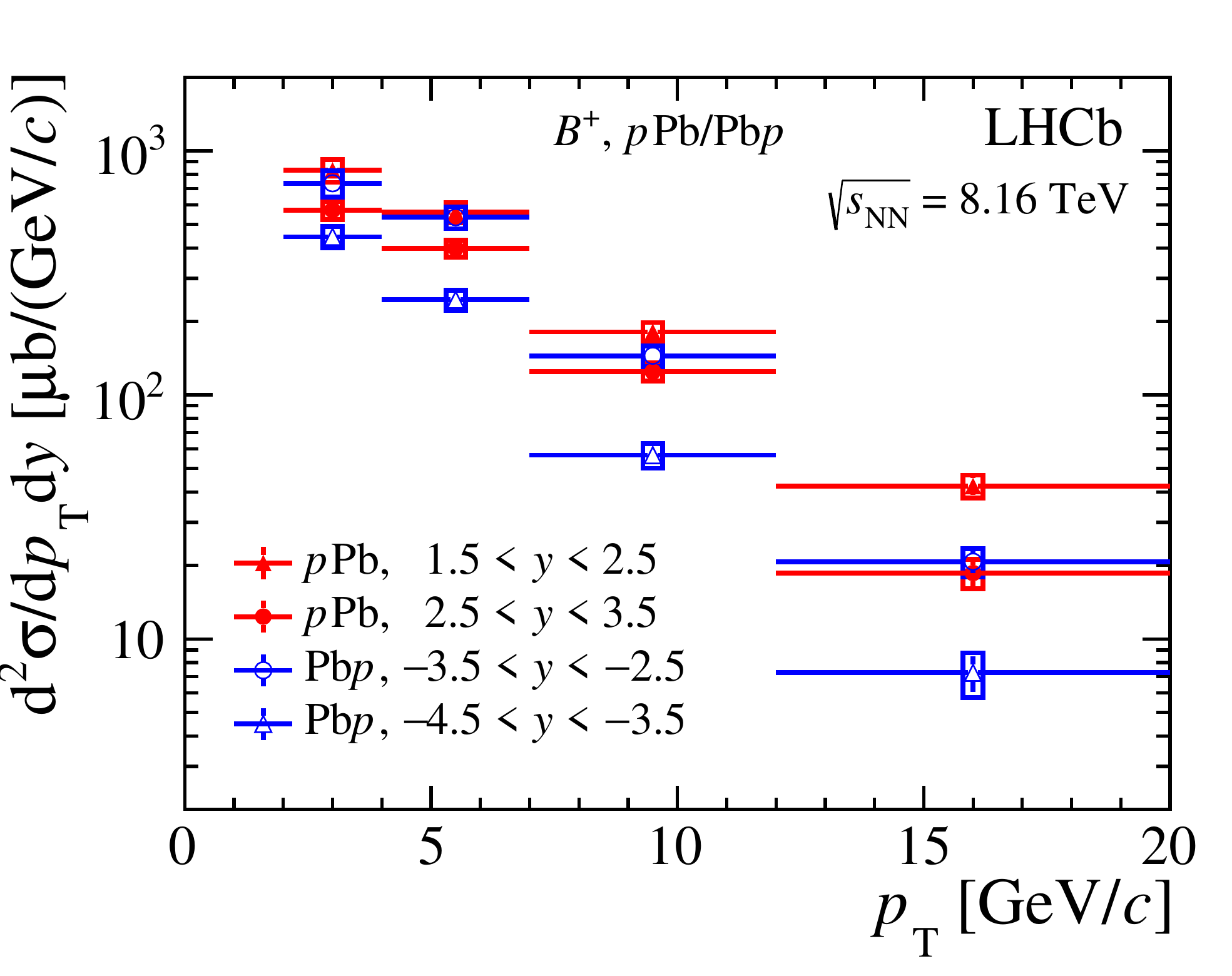}
  \includegraphics[width=0.49\textwidth]{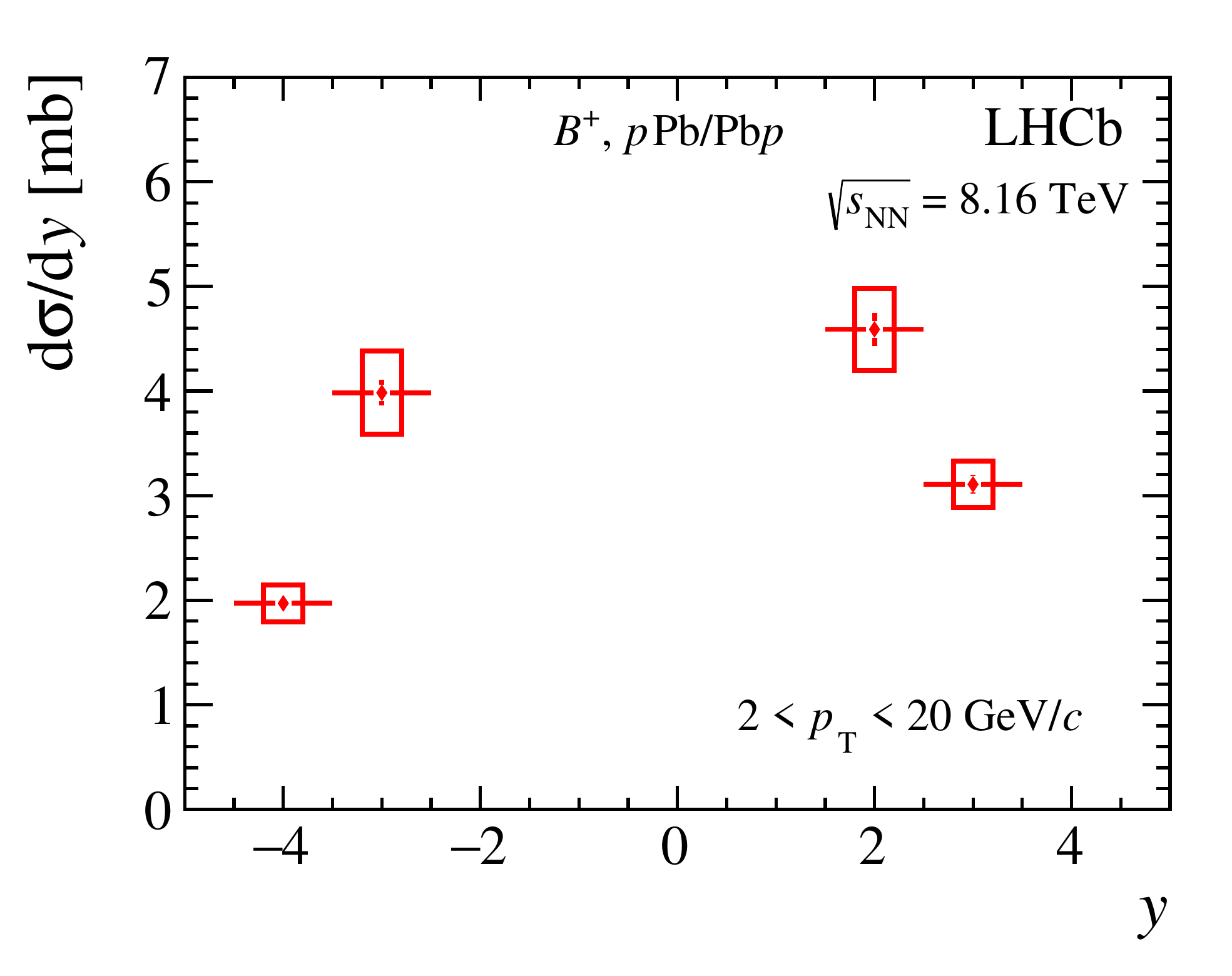}
  \caption{Production cross-section of $\Bp$ mesons as a function of (left) \pt in $y$ bins and (right) $y$ integrated over \pt.
    The vertical bars (boxes) show statistical (total) uncertainties.
    }
  \label{fig:Bplusxsec}
  \end{figure}

\begin{figure}[!tpb]
\centering

  \includegraphics[width=0.49\textwidth]{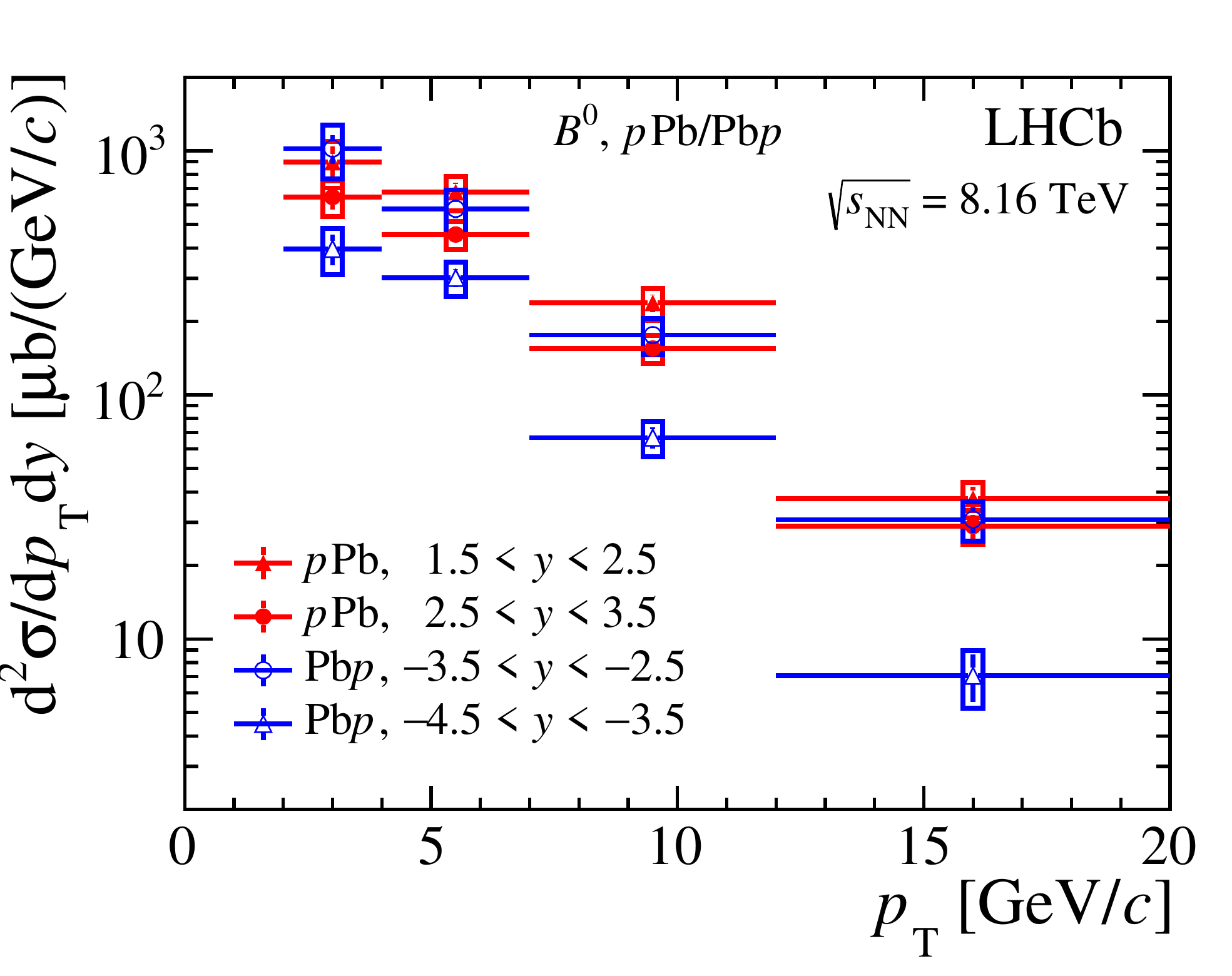}
  \includegraphics[width=0.49\textwidth]{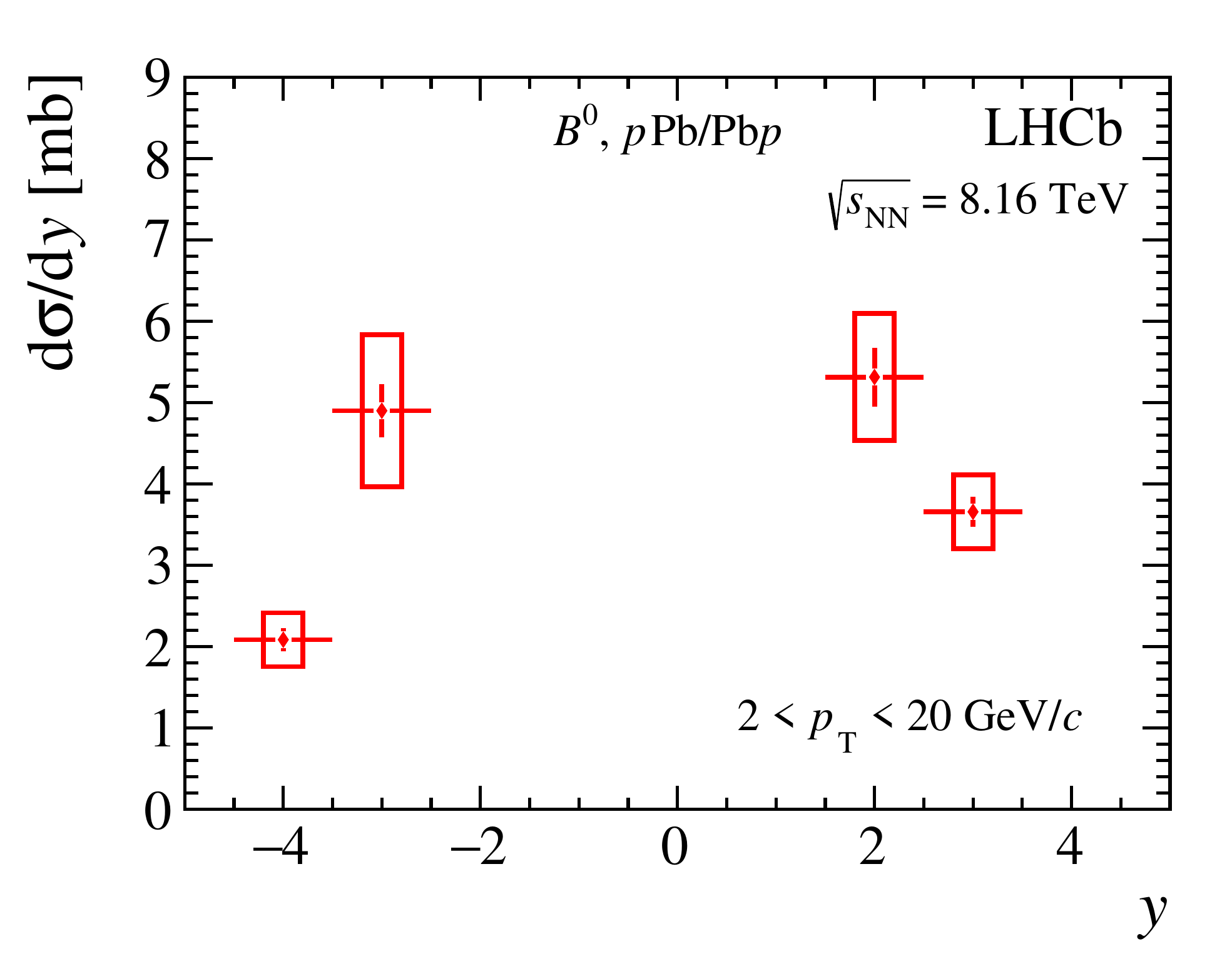}
  \caption{Production cross-section of $\Bz$ mesons as function of (left) \pt in $y$ bins and (right) $y$ integrated over \pt.
    The vertical bars (boxes) show statistical (total) uncertainties.
    }
  \label{fig:B0xsec}
  \end{figure}

The $\Bp$ cross-sections measured in the $\jpsi\Kp$ and $\Dzb\pip$ decay modes are consistent and their weighted average is reported. 
The weights are calculated using the statistical uncertainties combined with the systematic uncertainty due to the limited
sample size of the simulation samples.
The systematic uncertainties due to luminosity, kinematics, track reconstruction efficiency and kaon PID efficiency are entirely or strongly 
correlated, while those due to simulation sample size, muon and pion PID efficiencies, trigger selection and branching fractions 
are uncorrelated between the two decay modes.
The double-differential cross-section of the averaged $\Bp$ production in four rapidity bins as a function of \pt and integrated over
$\pt$ as a function of rapidity are shown in Fig.~\ref{fig:Bplusxsec} and reported in Table~\ref{tab:cross_section}.
The same quantities for \Bz production are displayed in Fig.~\ref{fig:B0xsec} and listed in
Table~\ref{tab:cross_section}. The measured cross-sections increase  towards central rapidity both at positive and at
negative rapidity.  A good precision is achieved in the $\Bp$ sample due to the averaging over two decay channels,
which allows for improved precision with respect to the measurement in each single $\Bp$ decay mode.

\begin{table}[!t]
\caption{
    Differential cross-sections of \Bp, \Bz and \Lb production in bins of \pt and $y$, $\frac{{\rm d}^2\sigma}{{\rm d}\pt {\rm d}y}(\mub/[\!\gevc])$, and in bins of $y$ integrated over $2<\pt<20\gevc$, $\frac{{\rm d}\sigma}{{\rm d}y} (\mub)$. The first uncertainty is
        statistical  and the second systematic.
   }
\centering
\scalebox{0.85}{
\begin{tabular}{lr@{$\,\pm\,$}r@{$\,\pm\,$}rr@{$\,\pm\,$}r@{$\,\pm\,$}rr@{$\,\pm\,$}r@{$\,\pm\,$}rr@{$\,\pm\,$}r@{$\,\pm\,$}r@{}}
    \toprule
$\pt\,\,(\mathrm{GeV}\!/c)$  & \multicolumn{3}{c}{$-4.5<y<-3.5$}  & \multicolumn{3}{c}{$-3.5<y<-2.5$}  &\multicolumn{3}{c}{$1.5<y<2.5$}&  \multicolumn{3}{c}{$2.5<y<3.5$}  \\
\midrule
\multicolumn{13}{c}{\Bp}\\
\midrule
$(\phantom{2}2,\phantom{2}4)$ & 441.1 & 25.8 & 36.0 & 735.7 & 45.6 & 78.7 &  831.1 & 54.8 & 69.8 & 571.3 & 30.8 & 36.6 \\
$(\phantom{1}4,\phantom{1}7)$ & 244.9 & 12.5 & 19.1 & 534.2 & 24.6 & 49.1 &        560.3 & 30.8 & 43.7 & 398.7 & 17.9 & 25.9 \\
$(\phantom{1}7,12)$ & 56.6 & 4.2 & 5.0 & 144.5 & 8.1 & 11.7 &        181.0 & 10.5 & 13.2 & 124.5 & 7.0 & 8.2 \\
$(12,20)$ & 7.3 & 1.2 & 0.9 & 20.7 & 2.1 & 1.7 & 42.3 & 3.5 & 3.0 &  18.6 & 2.2 & 1.3 \\
\midrule
$(\phantom{1}2,20)$ & 1971 & 69 & 162 & 3984 & 124 & 378 & 4590 & 156 & 358 &  3108 & 90 & 202 \\
\midrule
\multicolumn{13}{c}{\Bz} \\
\midrule
$(\phantom{1}2,\phantom{1}4)$ & 396.2 & 56.7 & 63.8 & 1020.8 & 136.8 & 213.3 &    898.0 & 144.6 & 130.2 & 645.9 & 70.4 & 81.4 \\
$(\phantom{1}4,\phantom{1}7)$ & 301.3 & 25.6 & 41.0 & 578.2 & 50.9 & 100.6 &  676.6 & 62.2 & 88.6 & 453.6 & 32.2 & 50.8 \\
$(\phantom{1}7,12)$ & 66.8 & 6.6 & 8.7 & 175.7 & 14.2 & 26.0 & 237.8 & 19.7 & 29.7 & 154.8 & 11.1 & 16.9 \\
$(12,20)$ & 7.1 & 1.6 & 1.0 & 30.8 & 3.7 & 4.3 & 37.5 & 4.4 & 4.4 & 29.0 & 3.3 & 3.2 \\
\midrule
$(\phantom{1}2,20)$ & 2086 & 142 & 298 & 4890 & 323 & 875 & 5332 & 357 & 693 &  3658 & 183 & 417 \\
\midrule
\multicolumn{13}{c}{\Lb}\\
\midrule
$(\phantom{1}2,\phantom{1}4)$ & 196.3 & 35.7 & 33.4 & 242.1 & 84.0 & 51.1 & 441.2 & 102.4 & 80.7 & 276.1 & 43.6 & 39.5 \\
$(\phantom{1}4,\phantom{1}7)$ & 106.8 & 14.9 & 16.8 & 244.6 & 33.7 & 43.3 & 289.5 & 40.8 & 44.6 & 219.7 & 21.1 & 29.0 \\
$(\phantom{1}7,12)$ & 35.7 & 4.4 & 5.4 & 85.6 & 9.2 & 13.6 & 107.5 & 11.9 & 14.7 &  48.7 & 5.7 & 6.4 \\
$(12,20)$ & 1.6 & 0.6 & 0.2 & 6.7 & 1.4 & 1.1 & 8.3 & 1.9 & 1.1 & 5.9 & 1.4 & 0.8 \\
\midrule
$(\phantom{1}2,20)$ & 935 & 91 & 149 & 1658 & 194 & 293 & 2305 & 244 & 360 &1480 & 111 & 198 \\
\bottomrule
\end{tabular}
}
\label{tab:cross_section}
\end{table}

\begin{figure}[!tpb]
\centering

  \includegraphics[width=0.49\textwidth]{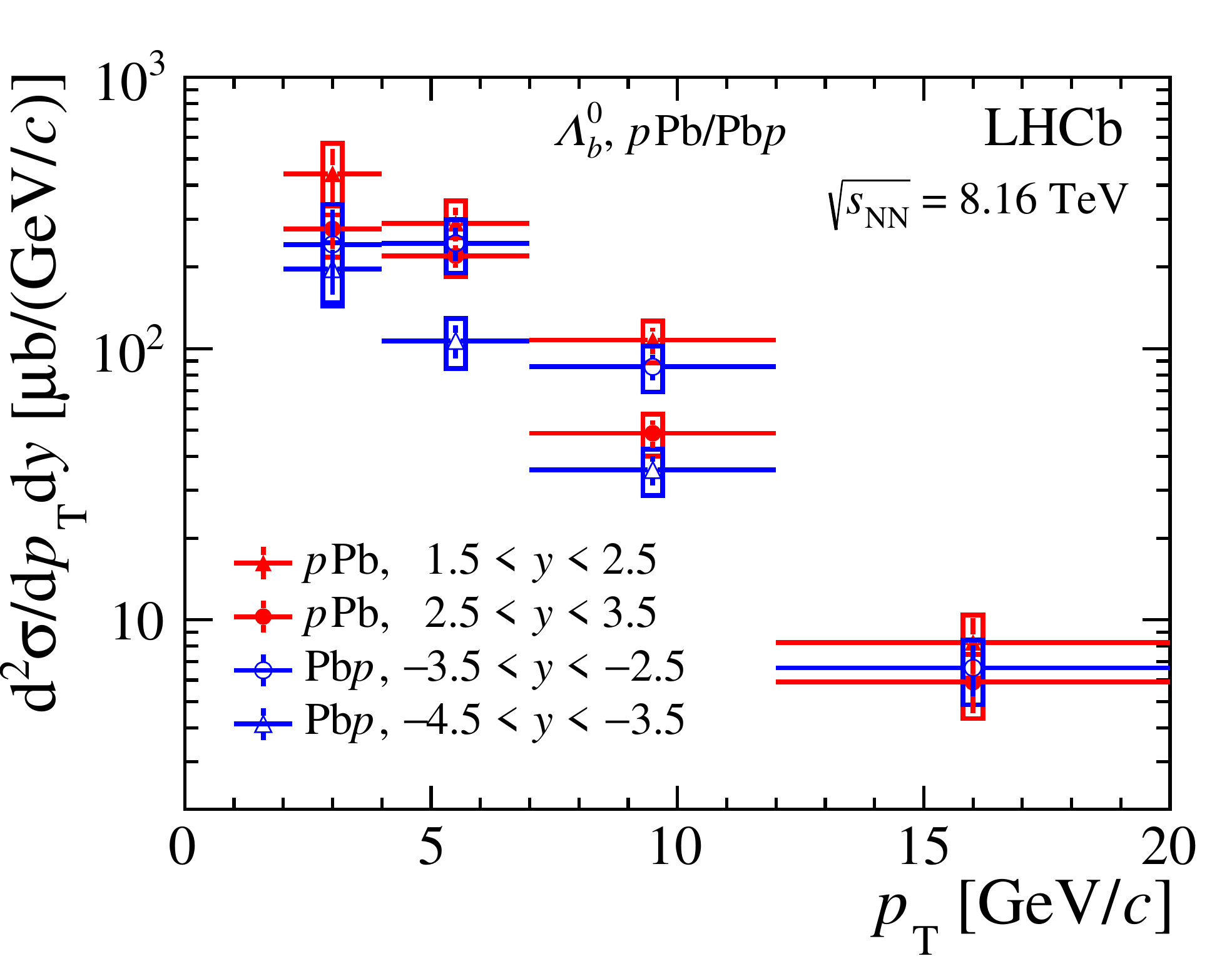}
  \includegraphics[width=0.49\textwidth]{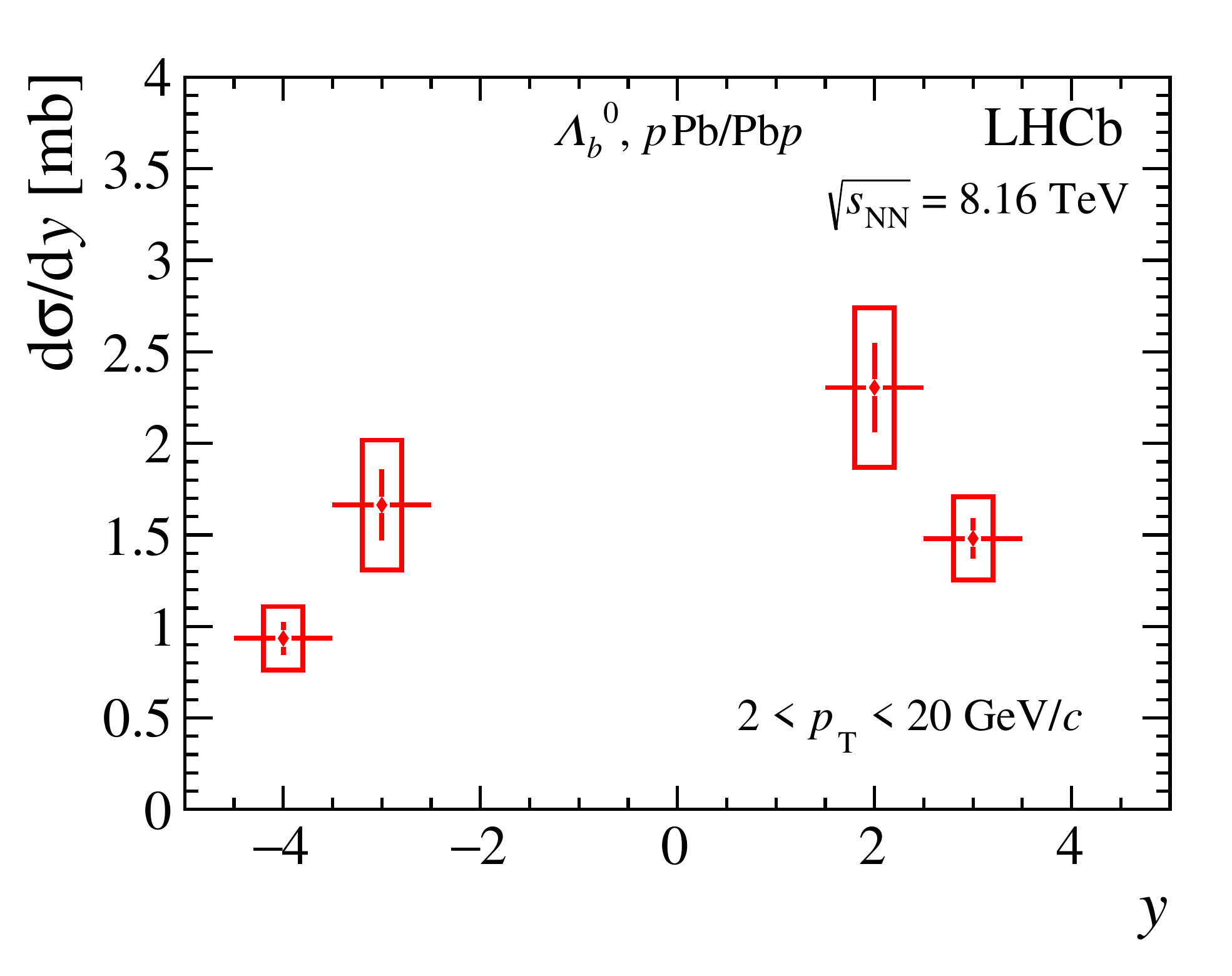}
  \caption{Production cross-section of $\Lb$ baryons as a function of (left) \pt in $y$ bins and (right) $y$ integrated over \pt.
    The vertical bars (boxes) show statistical (total) uncertainties.
    }
  \label{fig:Lbxsec}
\end{figure}

The double-differential cross-section of $\Lb$ production is shown in Fig.~\ref{fig:Lbxsec} in four rapidity bins as
a function of \pt and integrated over \pt as a function of rapidity, and is listed in Table~\ref{tab:cross_section}. 
The trend observed as a function of the two variables is similar to that of the $B$ mesons.

\begin{figure}[!tpb]
\centering
  \includegraphics[width=0.49\textwidth]{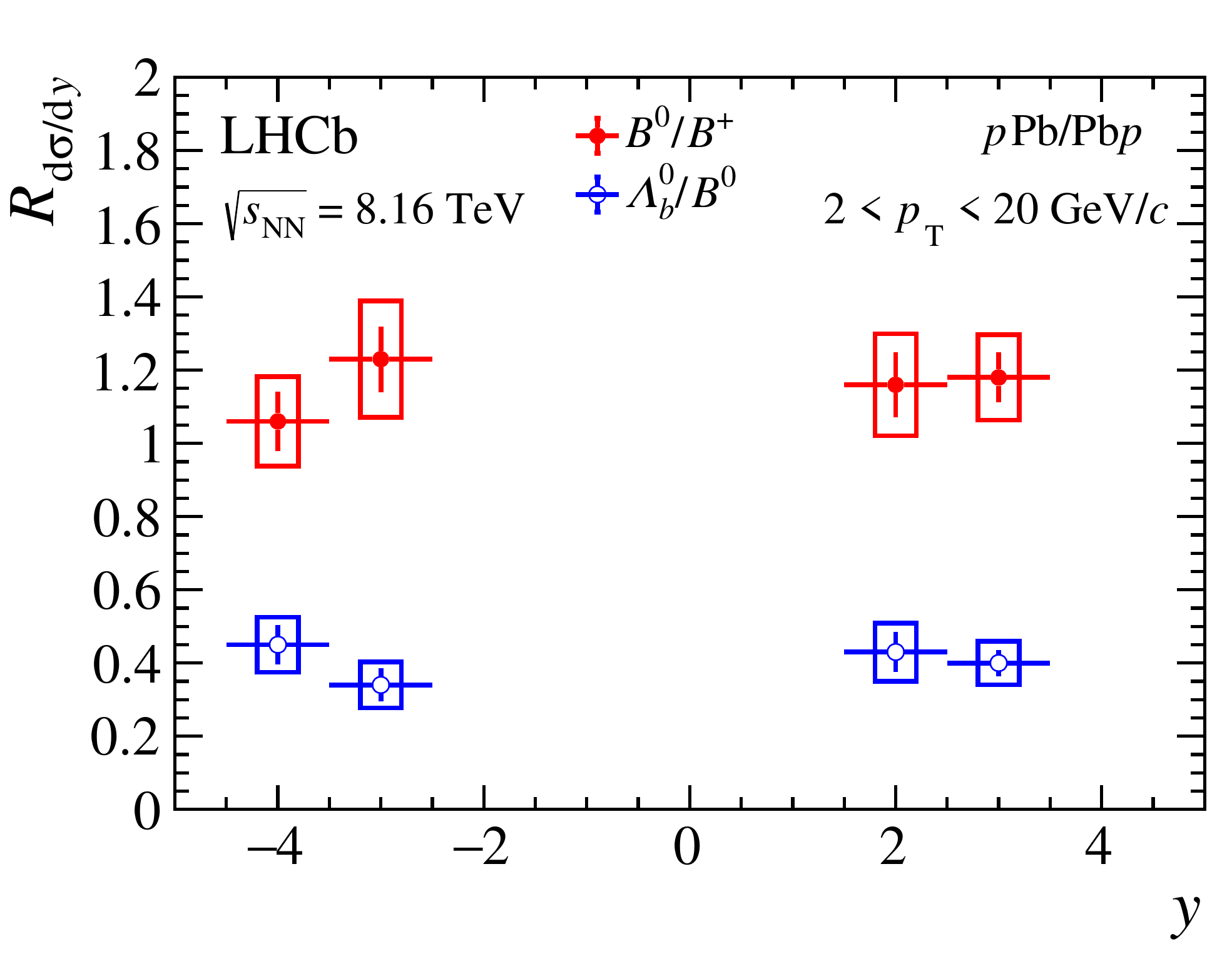}\\
  \includegraphics[width=0.49\textwidth]{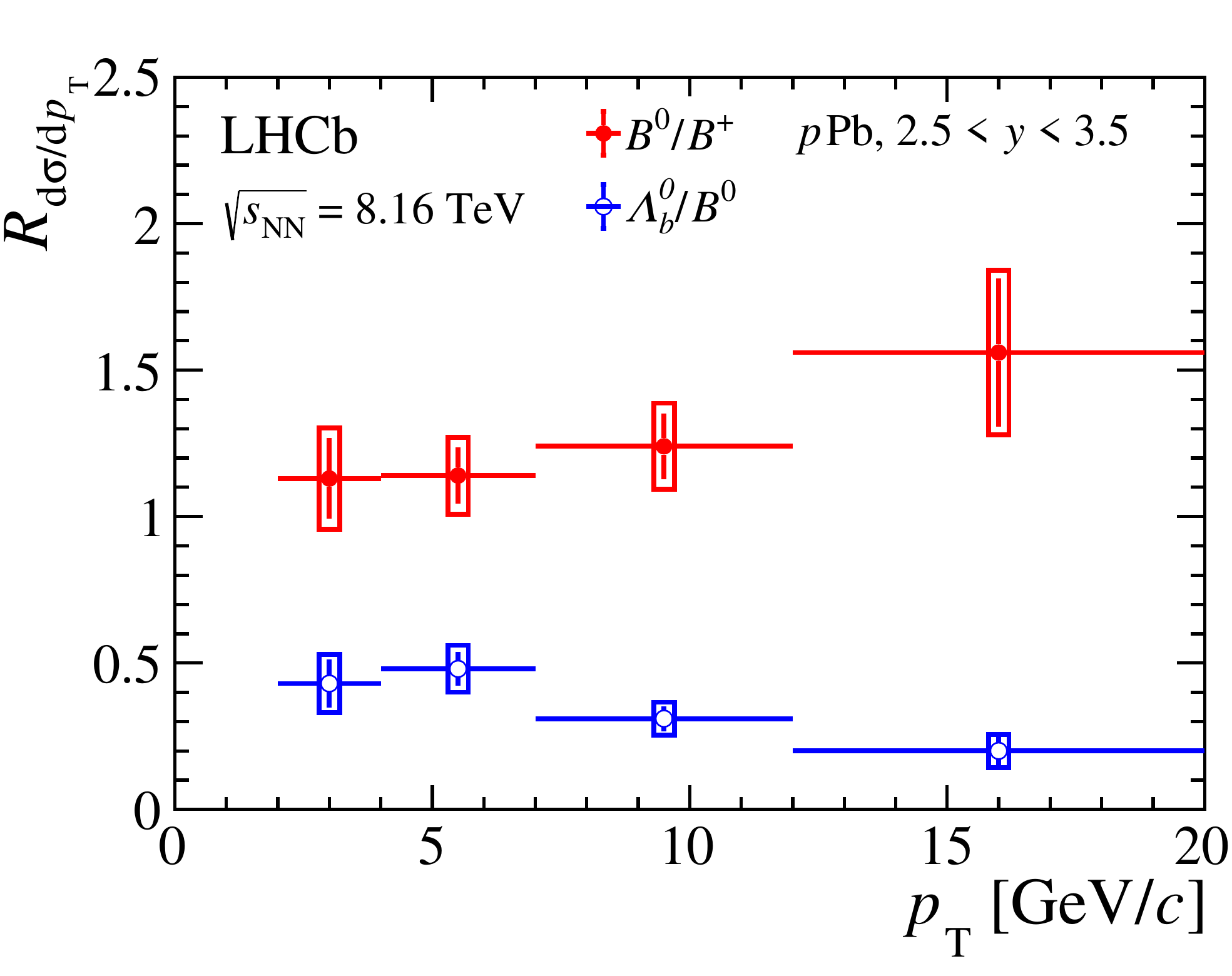}
  \includegraphics[width=0.49\textwidth]{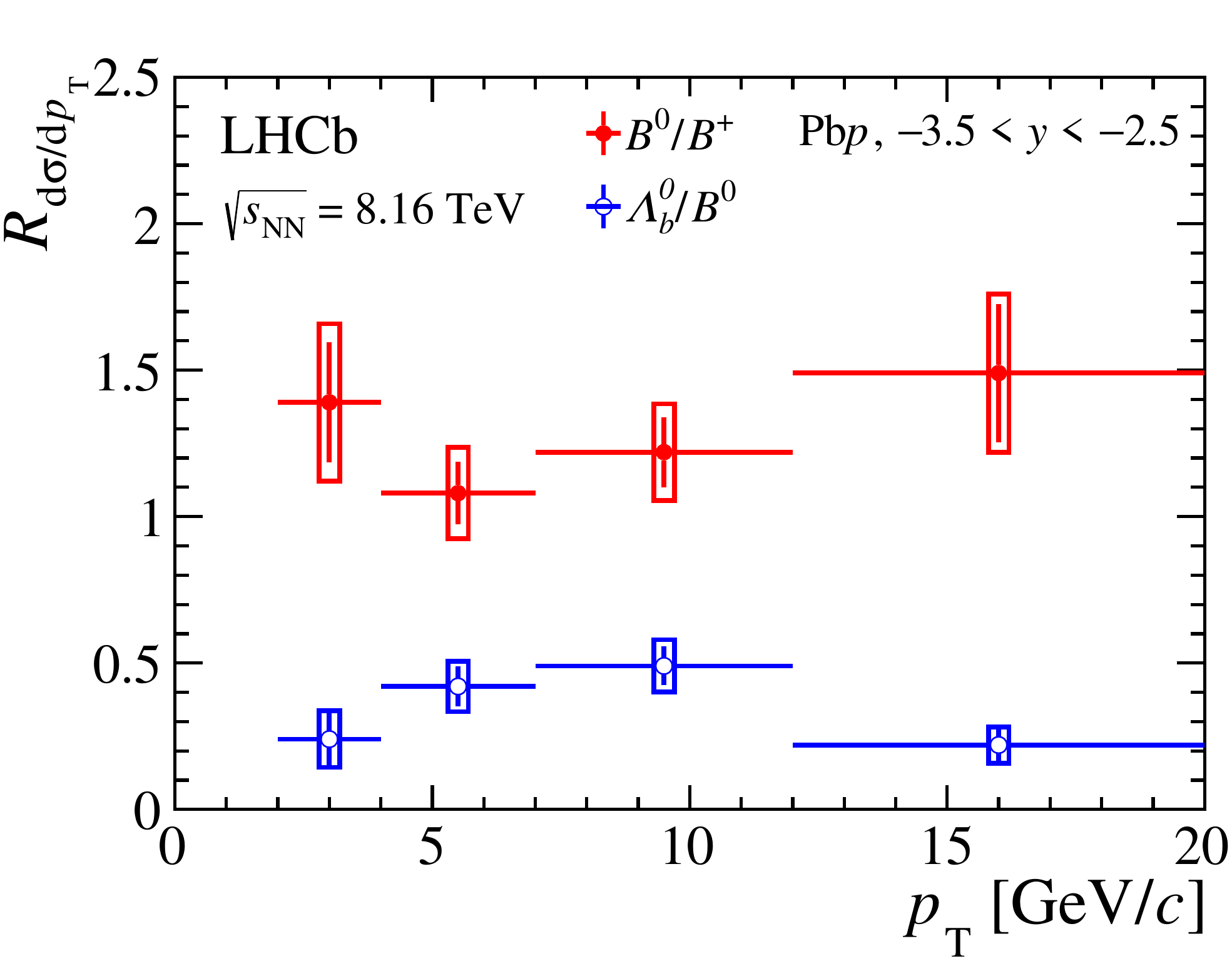}
  \caption{Production cross-section ratios of $\Lb$ baryons over $\Bz$ mesons and of $\Bz$ mesons over $\Bp$ mesons
    (top) as a function of $y$ integrated over \pt and as a function of \pt for (bottom left) $2.5<y<3.5$ 
    and (bottom right) $-3.5<y<-2.5$. The vertical bars (boxes) show statistical (total) uncertainties.
    }
  \label{fig:LboverB}
\end{figure}

In order to probe the hadronization in proton-lead collisions, ratios of \Bz over \Bp and \Lb over \Bz production cross-sections are
studied with results shown in Fig.~\ref{fig:LboverB}. Both ratios  show no significant rapidity dependence within 
experimental uncertainties. The ratio between meson species is consistent with being independent of $\ystar$ and 
\pt of the beauty hadrons. Most interestingly, the baryon-to-meson ratio shows a $\pt$ dependence with a significantly lower
value at the highest \pt compared to the \pt-integrated measurement. However, the current uncertainties do not allow to draw
firm conclusions. The production ratio, averaged over the kinematic range in the analysis, is measured to be
$0.41\pm0.06$ ($0.39\pm0.05$) for the \pPb (\Pbp) sample. The value is consistent with that measured by the LHCb
collaboration in $pp$ collisions~\cite{LHCb-PAPER-2012-037,LHCb-PAPER-2014-004,LHCb-PAPER-2011-018,Zhang:2017pmx}.

The cross-sections are used to calculate forward-backward ratios and nuclear modification factors. In the following, the
experimental results on these nuclear modification observables are compared with calculations using the
HELAC-onia generator~\cite{Shao:2012iz,Shao:2015vga,Lansberg:2016deg} with two different nuclear parton distribution
function (nPDF) sets, nCTEQ15~\cite{Kovarik:2015cma} and EPPS16~\cite{Eskola:2016oht}. For these calculations, the model
parameters are tuned to reproduce $pp$ cross-section measurements at the LHC. The uncertainties reflect those from
the corresponding nPDF parameterizations, and correspond to a 68\% confidence interval. A weighting of the current nPDF sets with heavy-flavor
measurements at the LHC was performed~\cite{Kusina:2017gkz} under the assumption that the modification of the nPDF is the main mechanism of
nuclear modification of heavy-flavor production. 
The corresponding predictions are shown together 
with their uncertainty bands under the label EPPS16$^*$~\cite{Kusina:2017gkz}. In the HELAC-onia framework, the nuclear matter effects are similar 
for the $\Bp$, $\Bz$ and $\Lb$ hadrons, \ie those possibly affecting the $b$-quark hadronization are not included. For this
reason, in the following the predictions are only compared with $\Bp$ production.

\subsection{Forward-backward ratios}

\begin{table}[!t]
\caption{
    Forward-backward ratios, $R_{\rm FB}$, of \Bp, \Bz and \Lb production in bins of \pt and integrated over $2.5<|y|<3.5$. The first uncertainty is
        statistical  and the second systematic.
   }
\centering
\scalebox{1.}{
\begin{tabular}{@{}lr@{$\,\pm\,$}r@{$\,\pm\,$}rr@{$\,\pm\,$}r@{$\,\pm\,$}rr@{$\,\pm\,$}r@{$\,\pm\,$}r@{}}
    \toprule
$\pt\,\,(\mathrm{GeV}\!/c)$  & \multicolumn{3}{c}{\Bp}  &\multicolumn{3}{c}{\Bz}&  \multicolumn{3}{c}{\Lb}  \\
\midrule
                     $(\phantom{1}2,\phantom{1}4)$ &          0.78 & 0.06 & 0.08 &          0.63 & 0.11 & 0.12 & 1.14 & 0.43 & 0.20 \\
                     $(\phantom{1}4,\phantom{1}7)$ &          0.75 & 0.05 & 0.06 &          0.78 & 0.09 & 0.12  & 0.90 & 0.15 & 0.13 \\
                    $(\phantom{1}7,12)$ &          0.86 & 0.07 & 0.06 &          0.88 & 0.10 & 0.10 & 0.57 & 0.09 & 0.06 \\
                  $(12,20)$  &          0.90 & 0.14 & 0.07 &          0.94 & 0.16 & 0.10 & 0.89 & 0.28 & 0.10 \\
\midrule
                   $(\phantom{1}2,20)$  &          0.78 & 0.03 & 0.07 &          0.75 & 0.06 & 0.12 & 0.89 & 0.12 & 0.12 \\
\bottomrule
\end{tabular}
}
\label{tab:rfb}
\end{table}

\begin{figure}[!tbp]
\centering
  \includegraphics[width=0.49\textwidth]{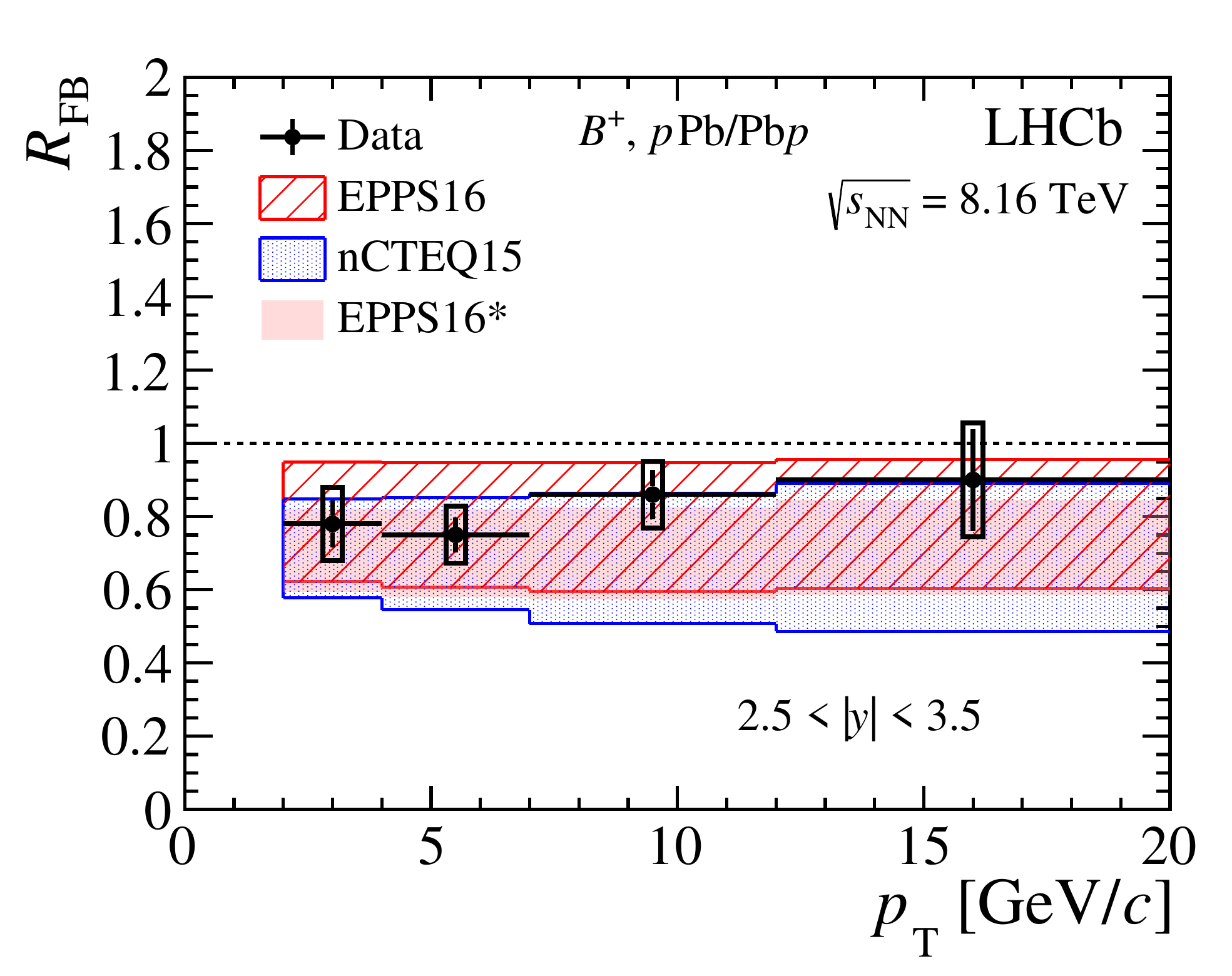}
  \includegraphics[width=0.49\textwidth]{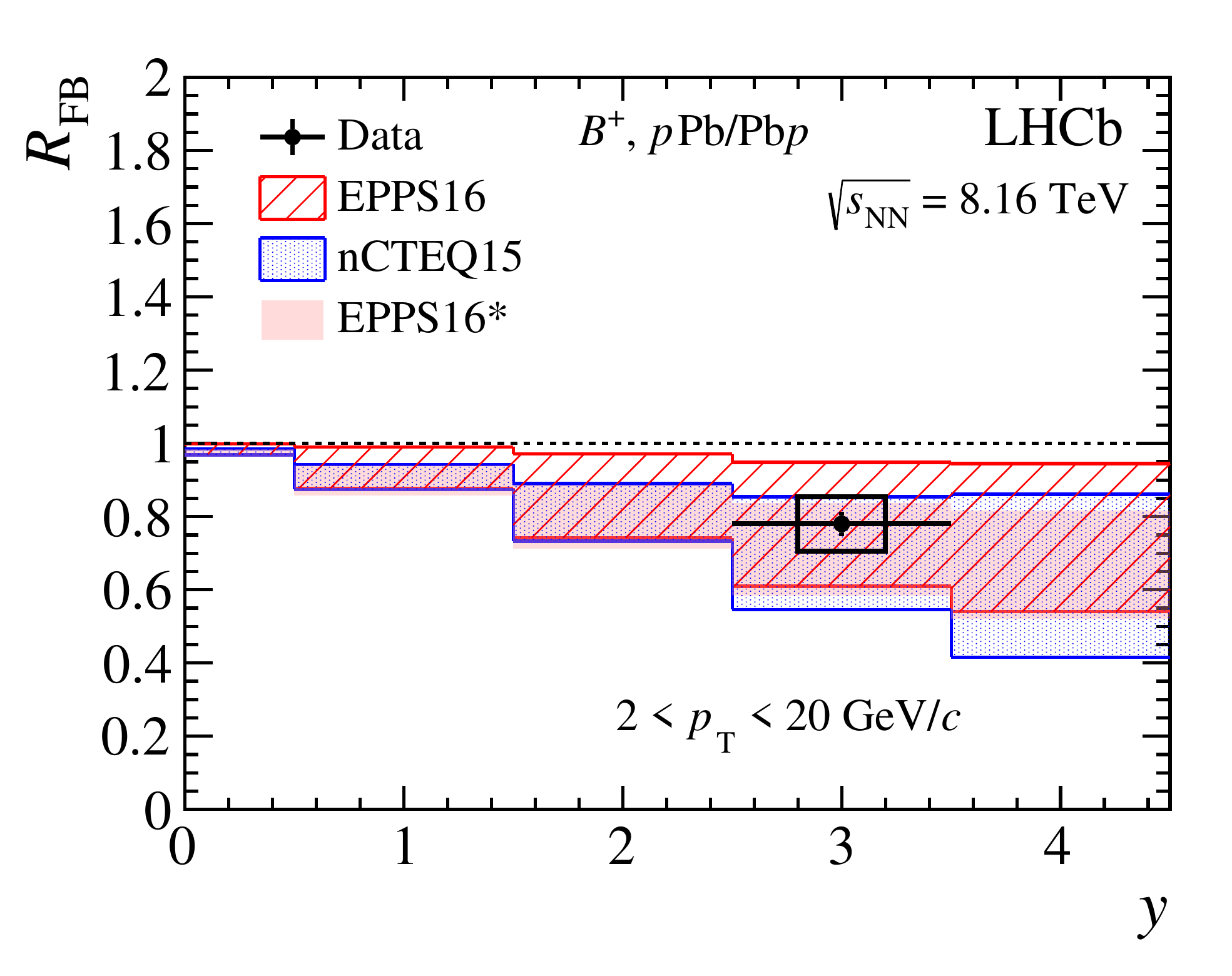}
  \caption{Forward-backward ratio, $R_{\rm FB}$, for $\Bp$ mesons as a function of (left) \pt and (right) $y$ in proton-lead collisions compared with HELAC-onia calculations using different nPDF sets.
    For the data points, the vertical bars (boxes) represent the statistical (total) uncertainties. The rapidity range
    $2.5<|\ystar|<3.5$ represents the common range between the \pPb and \Pbp data samples studied in this analysis.
    }
  \label{fig:RFB_Bplus}
  \end{figure}

The forward-backward production ratio of $\Bp$ mesons is shown in Fig.~\ref{fig:RFB_Bplus} as a function of \pt
and $y$, while the corresponding values are reported in Table~\ref{tab:rfb}. A significant
suppression of the production in the \pPb sample with respect to that in the \Pbp data is measured at the level of 20\% when integrating over \pt. Within the
experimental uncertainty, no dependence as a function of \pt is  observed. The HELAC-onia
calculations using EPPS16 and nCTEQ15 are in agreement with the experimental data. 
The EPPS16$^*$ set exhibits the smallest uncertainties and is also in agreement with data.

\begin{figure}[!tpb]
\centering
  \includegraphics[width=0.47\textwidth]{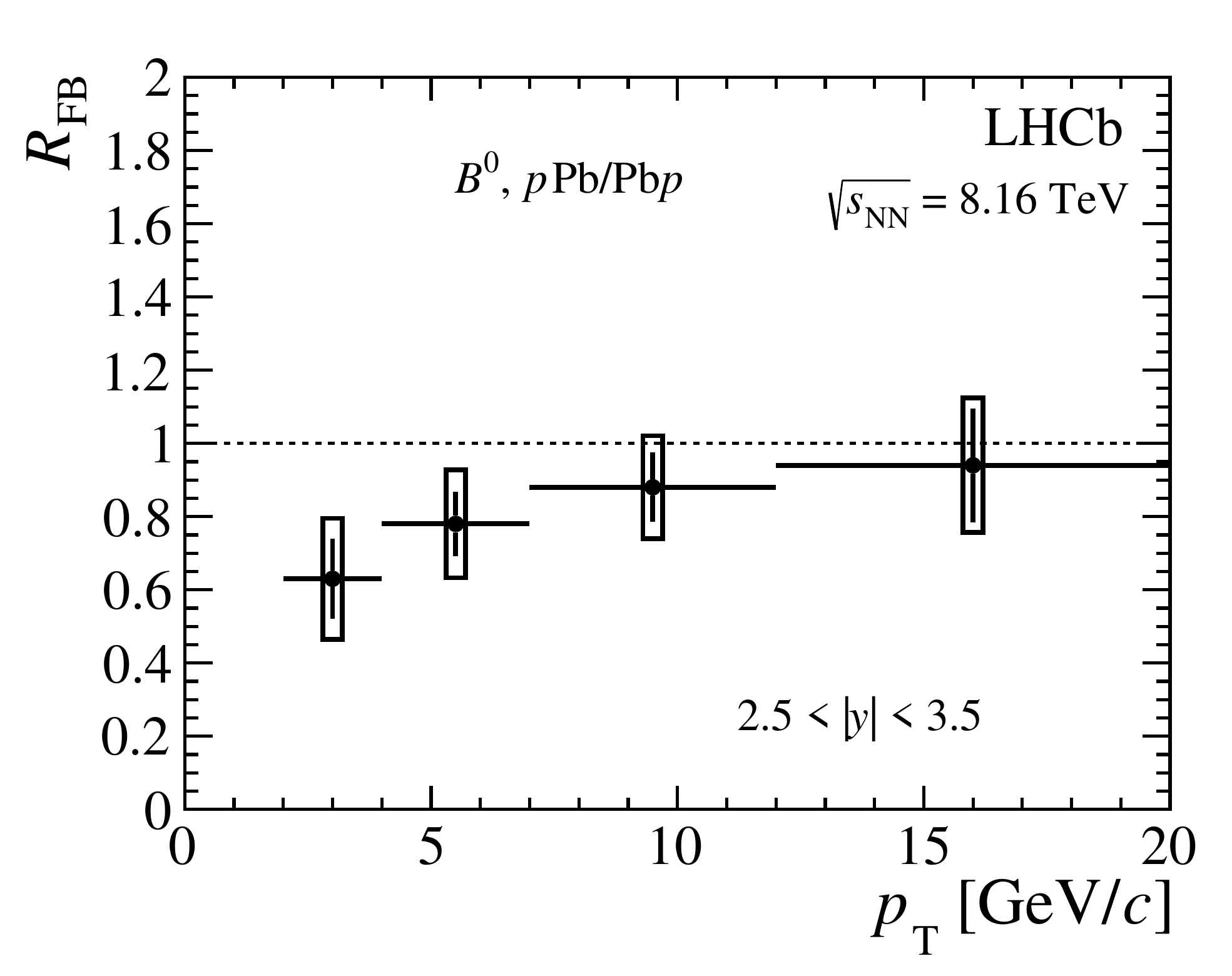}
  \includegraphics[width=0.47\textwidth]{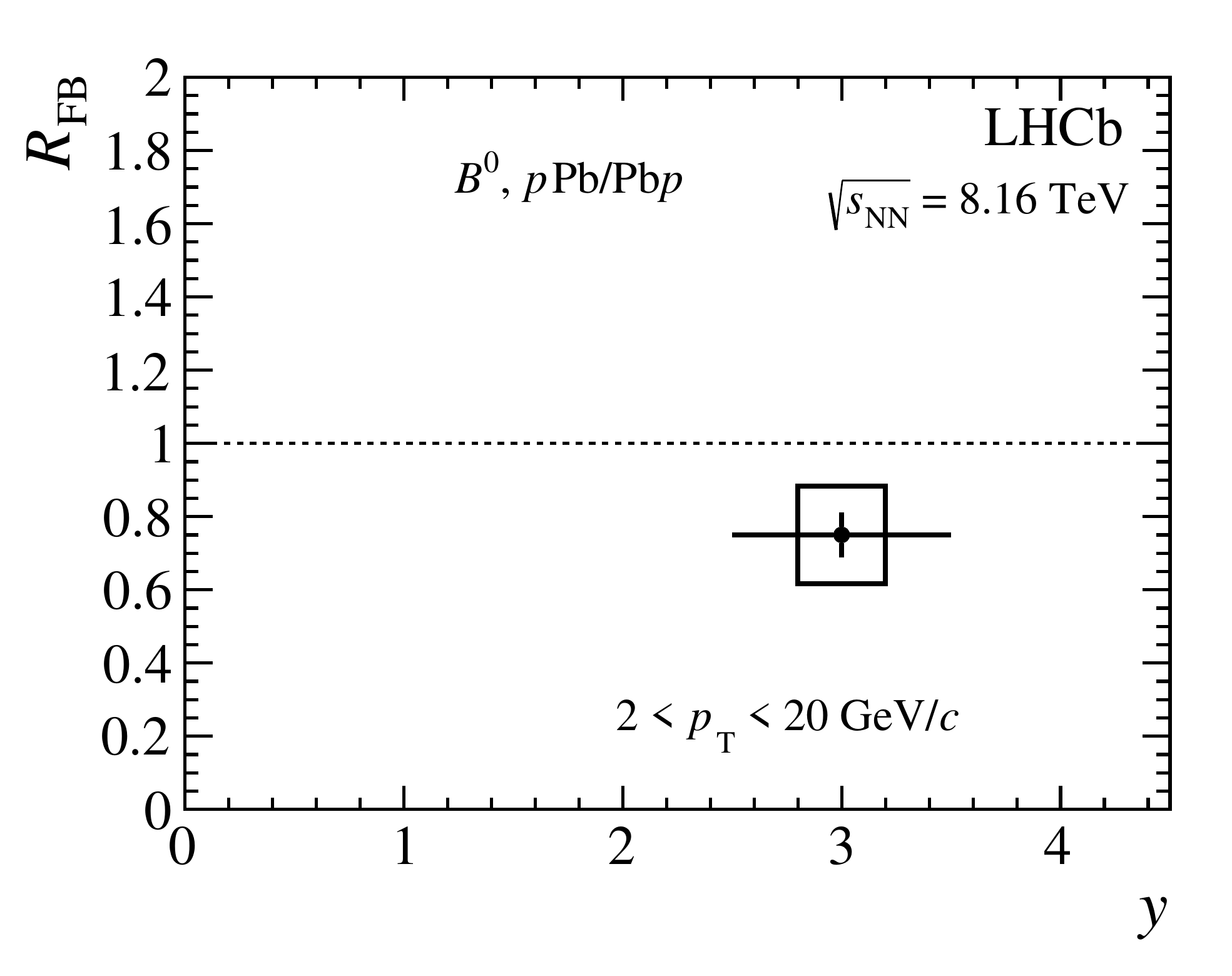}
  \caption{Forward-backward ratio, $R_{\rm FB}$, of $\Bz$ mesons as a function of (left) \pt and as a function of (right) $y$ in proton-lead collisions.
    The vertical bars (boxes) represent the statistical (total) uncertainties.
    }
  \label{fig:RFB_B0}
  \end{figure}

The $R_{\rm FB}$ ratio as a function of $\pt$ for $\Bz$ mesons and the \pt-integrated value is shown in
Fig.~\ref{fig:RFB_B0} and given in Table~\ref{tab:rfb}. A significant suppression is observed when integrating over the
considered \pt range, consistent with the value measured for $\Bp$ mesons. No significant dependence on \pt is seen
within the current experimental uncertainties. 

\begin{figure}[!tpb]
\centering
  \includegraphics[width=0.47\textwidth]{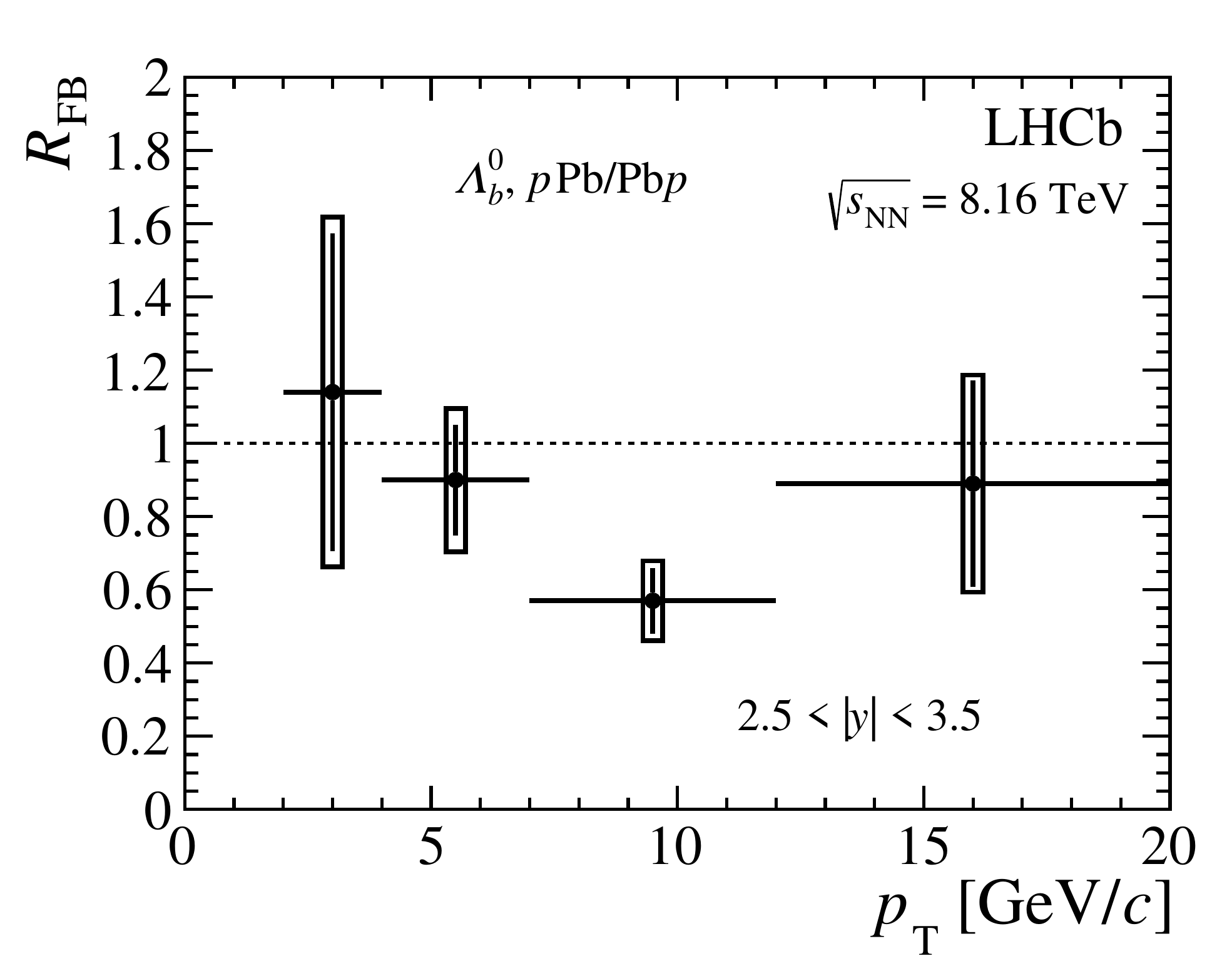}
  \includegraphics[width=0.47\textwidth]{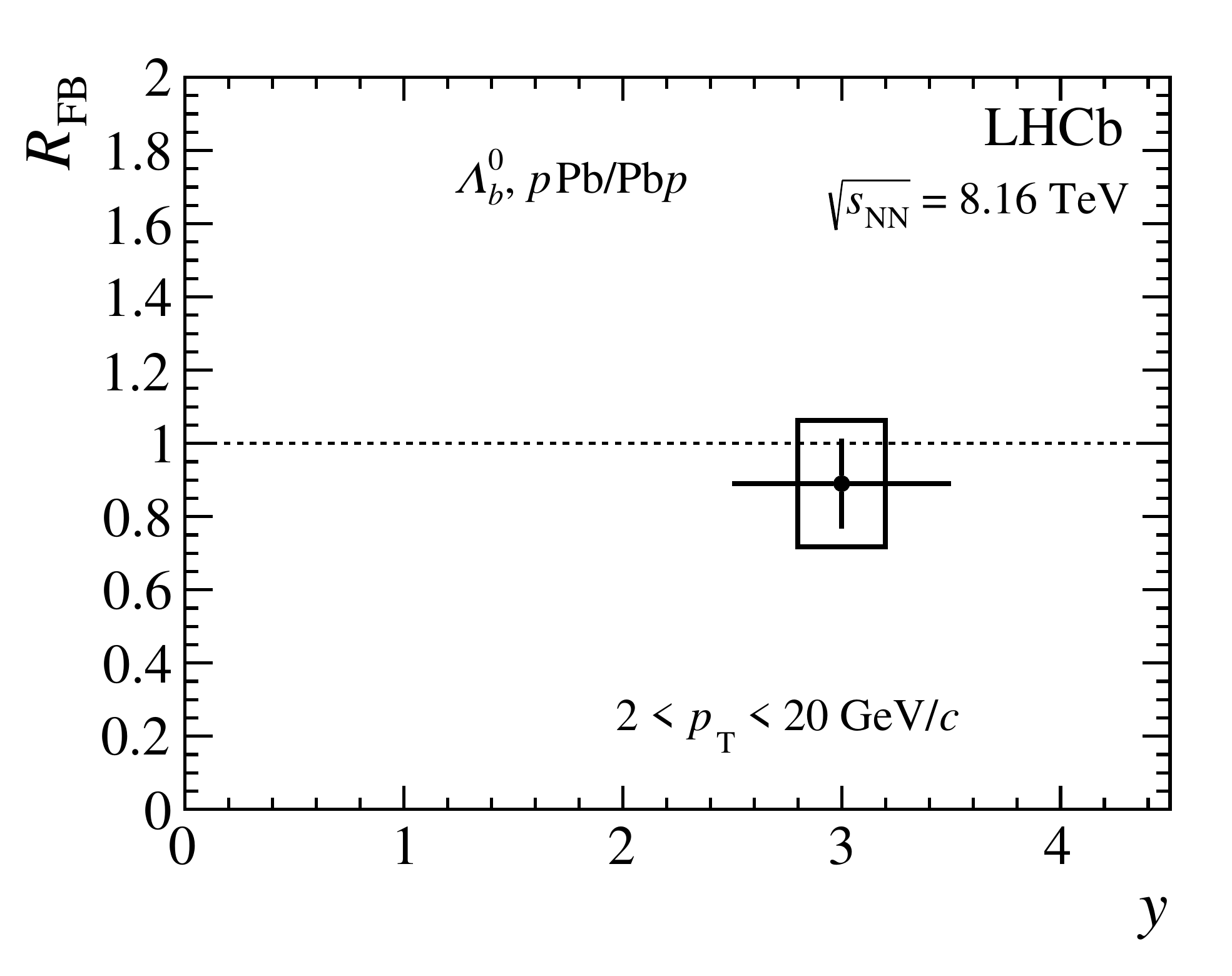}
  \caption{Forward-backward ratio, $R_{\rm FB}$, of $\Lb$ baryons as a function of (left) \pt and (right) $y$ in proton-lead collisions.
    The vertical bars (boxes) represent the statistical (total) uncertainties.
    }
  \label{fig:RFB_Lb}
\end{figure}

\begin{figure}[!tpb]
\centering
  \includegraphics[width=0.47\textwidth]{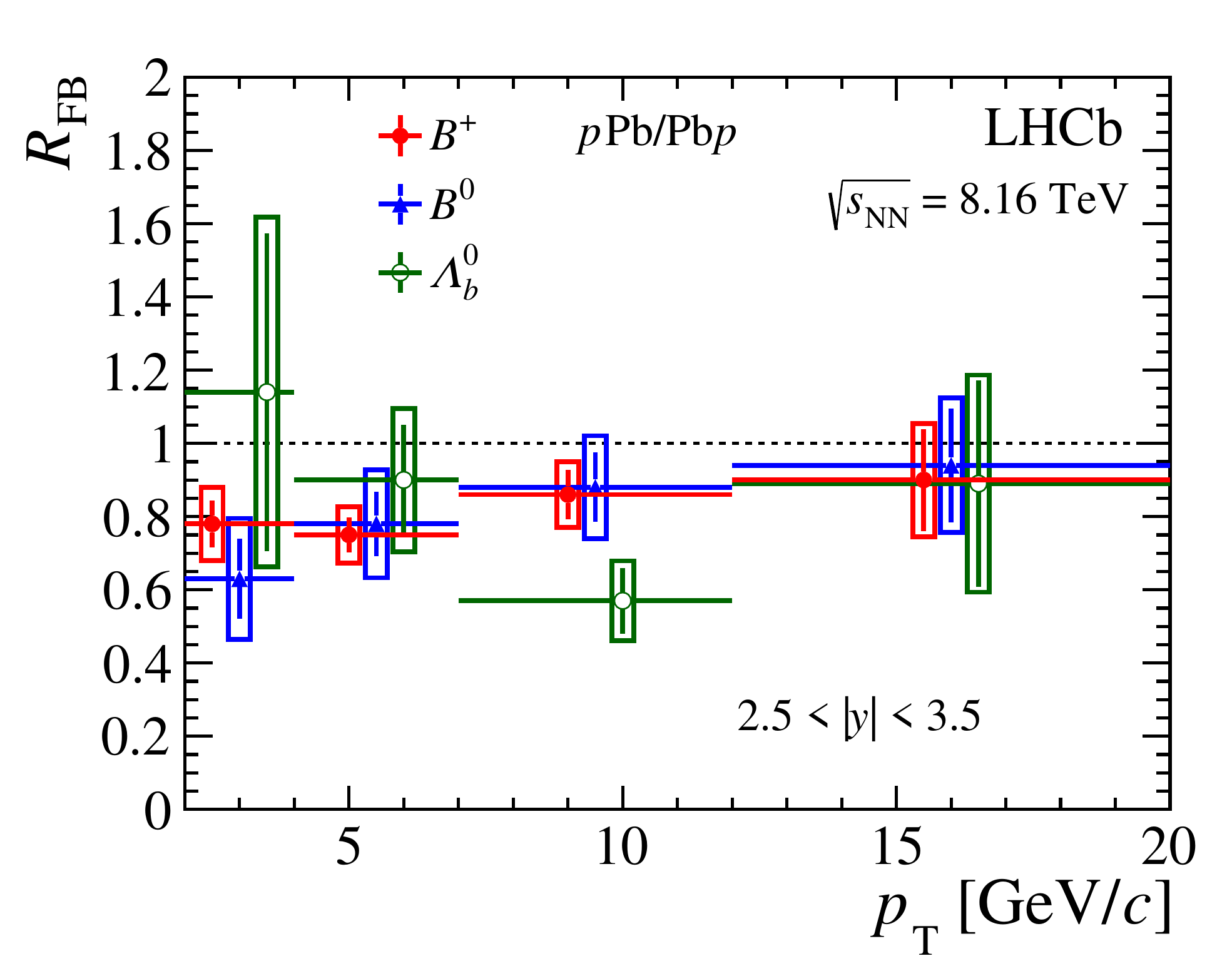}
  \includegraphics[width=0.47\textwidth]{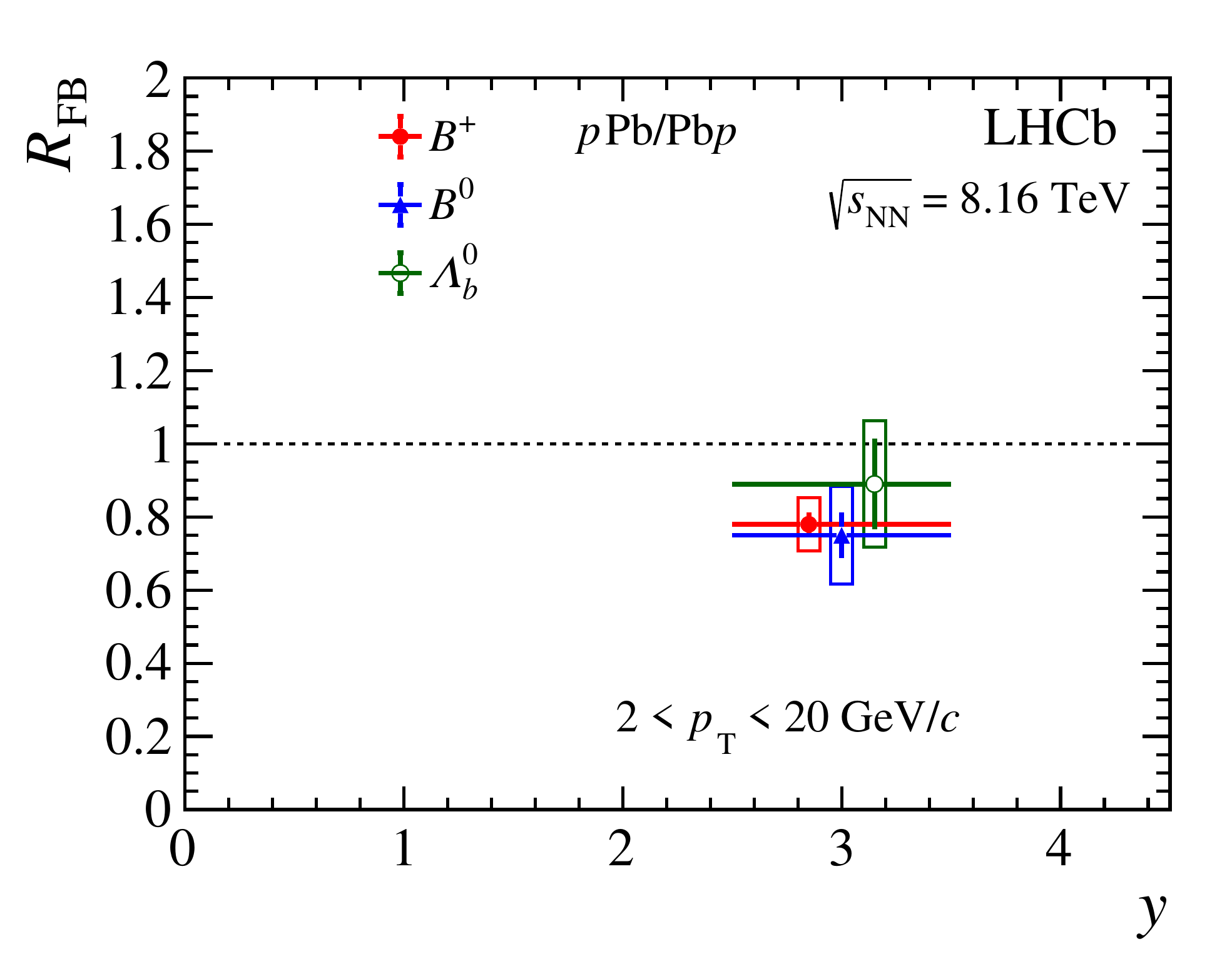}
    \caption{Forward-backward ratio, $R_{\rm FB}$, of (red) $\Bp$, (blue) $\Bz$ mesons and (green) $\Lb$ baryons as a function of (left) \pt and (right) $y$ in proton-lead collisions.
    The vertical bars (boxes) represent the statistical (total) uncertainties. Data points are shifted horizontally for better visibility. }
  \label{fig:RFB_comparison}
\end{figure}

In Fig.~\ref{fig:RFB_Lb}, the forward-backward cross-section ratio, $R_{\rm FB}$, of $\Lb$ production is shown. The
numerical values are summarized in Table~\ref{tab:rfb}. The observed central value of $R_{\rm FB}$ for the $\Lb$  baryon
is consistent with the measured value for the two $b$-meson species and with the no-suppression hypothesis. 
A significant suppression of $\Lb$ production in \pPb data compared to \Pbp data  is observed for the most precisely
measured bin, between 7 and 12\gevc. The $R_{\rm FB}$ measurement of $\Lb$ baryons is consistent with
the modifications observed for the beauty mesons within the uncertainties for all kinematic bins.
In Fig.~\ref{fig:RFB_comparison}, the values of $R_{\rm FB}$ as a function of \pt and as a function of $\ystar$ for the three
hadrons are compared directly.

\subsection{Nuclear modification factors}
\begin{figure}[!tpb]
\centering

  \includegraphics[width=0.49\textwidth]{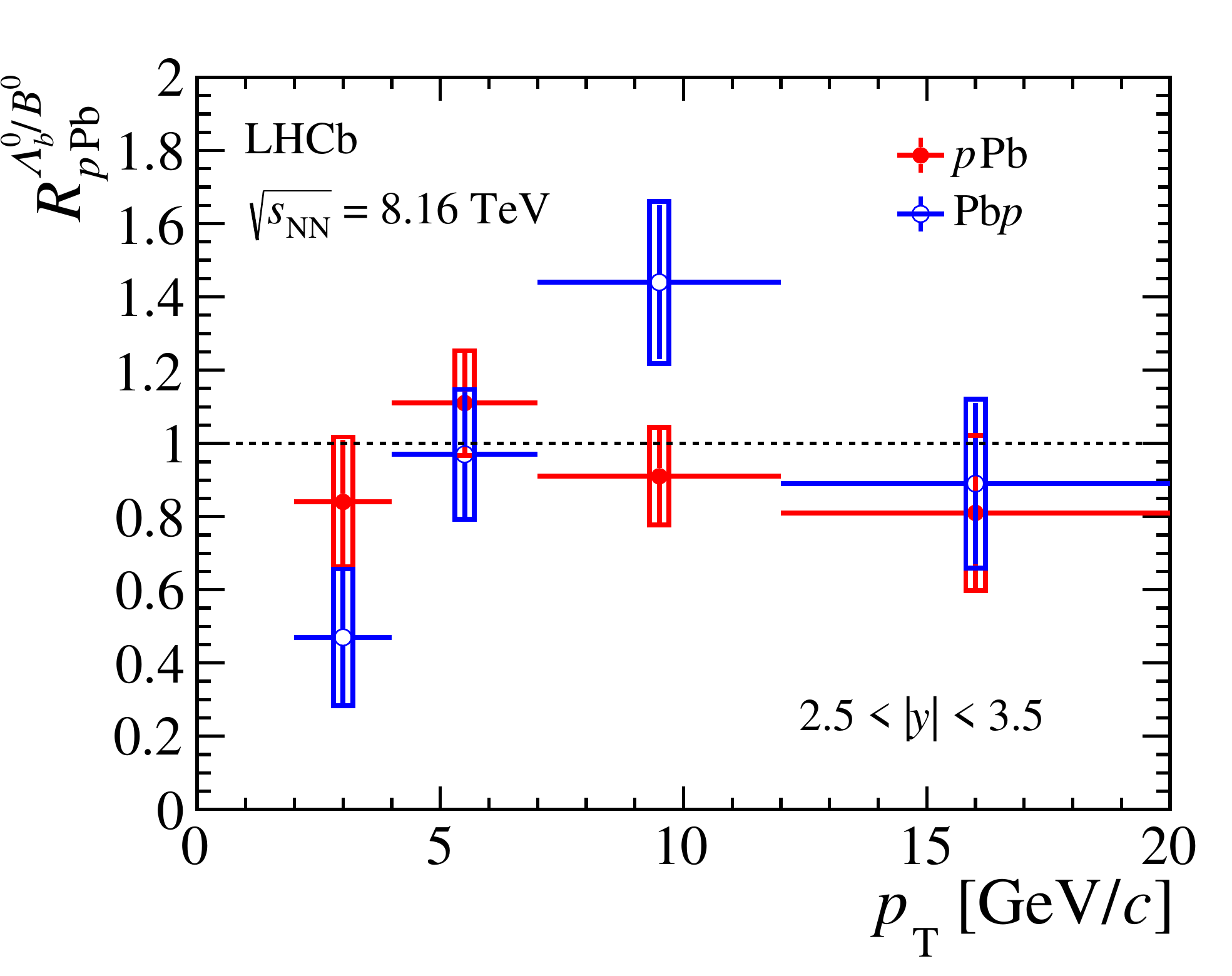}
  \includegraphics[width=0.49\textwidth]{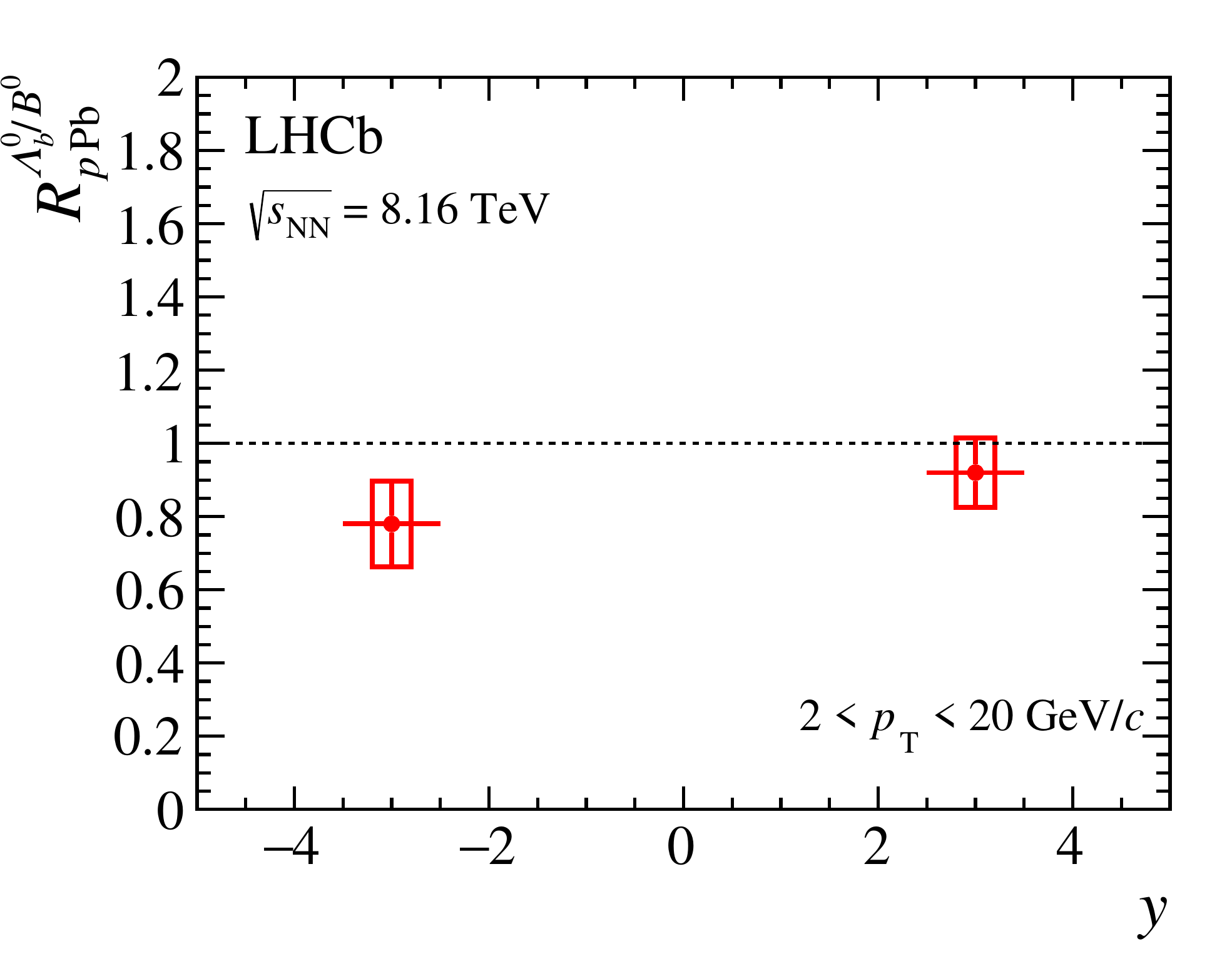}
  \caption{Ratio of nuclear modification factors, $R^{\Lb/\Bz}_{\pPb}$, as a function of (left) \pt and (right) $y$ in $\pPb$ and $\Pbp$ collisions.
    The vertical bars (boxes) represent the statistical (total) uncertainties.
    }
  \label{fig:doubleratio}
\end{figure}

In order to gain insight into potential modifications of the $b$-quark hadronization in $\pPb$ and $\Pbp$ collisions with respect to $pp$ collisions, 
the $\Lb/\Bz$ cross-section ratio shown in Fig.~\ref{fig:LboverB} is divided by the corresponding measurement in 
$pp$ collisions at $\sqrt{s}=7\tev$~\cite{LHCb-PAPER-2014-004}.
Neglecting the dependence on the collision energy of the hadronization with respect to the
experimental uncertainties,  the quantity corresponds to the ratios of nuclear
modification factors
\begin{equation}
    R^{\Lb/\Bz}_{\pPb}\equiv \frac{R^{\Lb}_{\pPb}}{R^{\Bz}_{\pPb}}.
    \end{equation}
If the overall nuclear effects for $\Bz$ mesons and $\Lb$ baryons are identical, $R^{\Lb/\Bz}_{\pPb}$ is expected to be unity. 
This double ratio is presented as a function of \pt and
\ystar in Fig.~\ref{fig:doubleratio} and in Table~\ref{tab:ratios}. At positive rapidity, the value of the ratio in all kinematic bins is consistent
with unity.  At negative rapidity (\Pbp), the lowest \pt bin exhibits a value smaller than one by more than two standard deviations and the third bin exceeds
one by about two standard deviations. The $\pt$-integrated value in the rapidity range $-3.5<y<-2.5$ is about two standard deviations away from unity.
However, more data are required to test whether there are different nuclear effects in beauty mesons and baryons. 
It would be interesting to check from the theory side whether deviations from unity are expected from models of quark recombination effects in
heavy-flavor production in heavy-ion collisions.

\begin{table}[!tpb]
\caption{
    Ratios of nuclear modification factors, $R^{\Lb/\Bz}_{\pPb}$, in bins of \pt and integrated over $2.5<|y|<3.5$, for $\pPb$ and $\Pbp$ samples.
    The first uncertainty is statistical and the second systematic.
   }
\centering
\scalebox{1.}{
\begin{tabular}{@{}lr@{$\,\pm\,$}r@{$\,\pm\,$}rr@{$\,\pm\,$}r@{$\,\pm\,$}r@{}}
    \toprule
$\pt\,\,(\mathrm{GeV}\!/c)$  & \multicolumn{3}{c}{$\pPb$}  &\multicolumn{3}{c}{$\Pbp$} \\
\midrule
                     $(\phantom{1}2,\phantom{1}4)$ &          0.84 & 0.17 & 0.05 &          0.47 & 0.18 & 0.05 \\
                     $(\phantom{1}4,\phantom{1}7)$ &          1.11 & 0.14 & 0.03 &          0.97 & 0.17 & 0.05  \\
                    $(\phantom{1}7,12)$ &          0.91 & 0.13 & 0.03 &          1.44 & 0.21 & 0.07 \\
                   $(12,20)$ &          0.81 & 0.21 & 0.03 &          0.89 & 0.22 & 0.07 \\
\midrule
                    $(\phantom{1}2, 20)$ &          0.92 & 0.09 & 0.03 &          0.78 & 0.11 & 0.04 \\
\bottomrule
\end{tabular}
}
\label{tab:ratios}
\end{table}

\begin{figure}[!tpb]
\centering
  \includegraphics[width=0.49\textwidth]{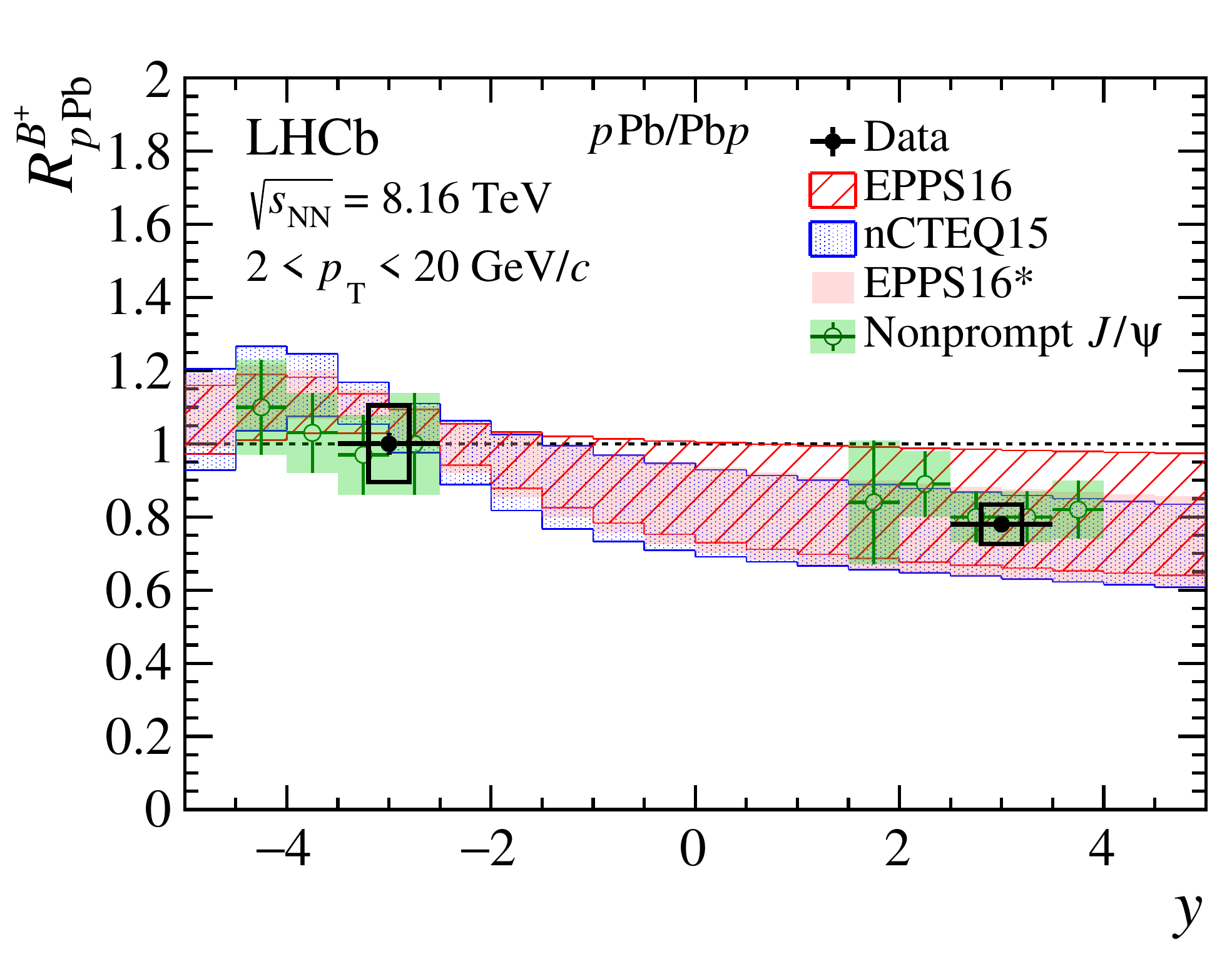}\\
  \includegraphics[width=0.49\textwidth]{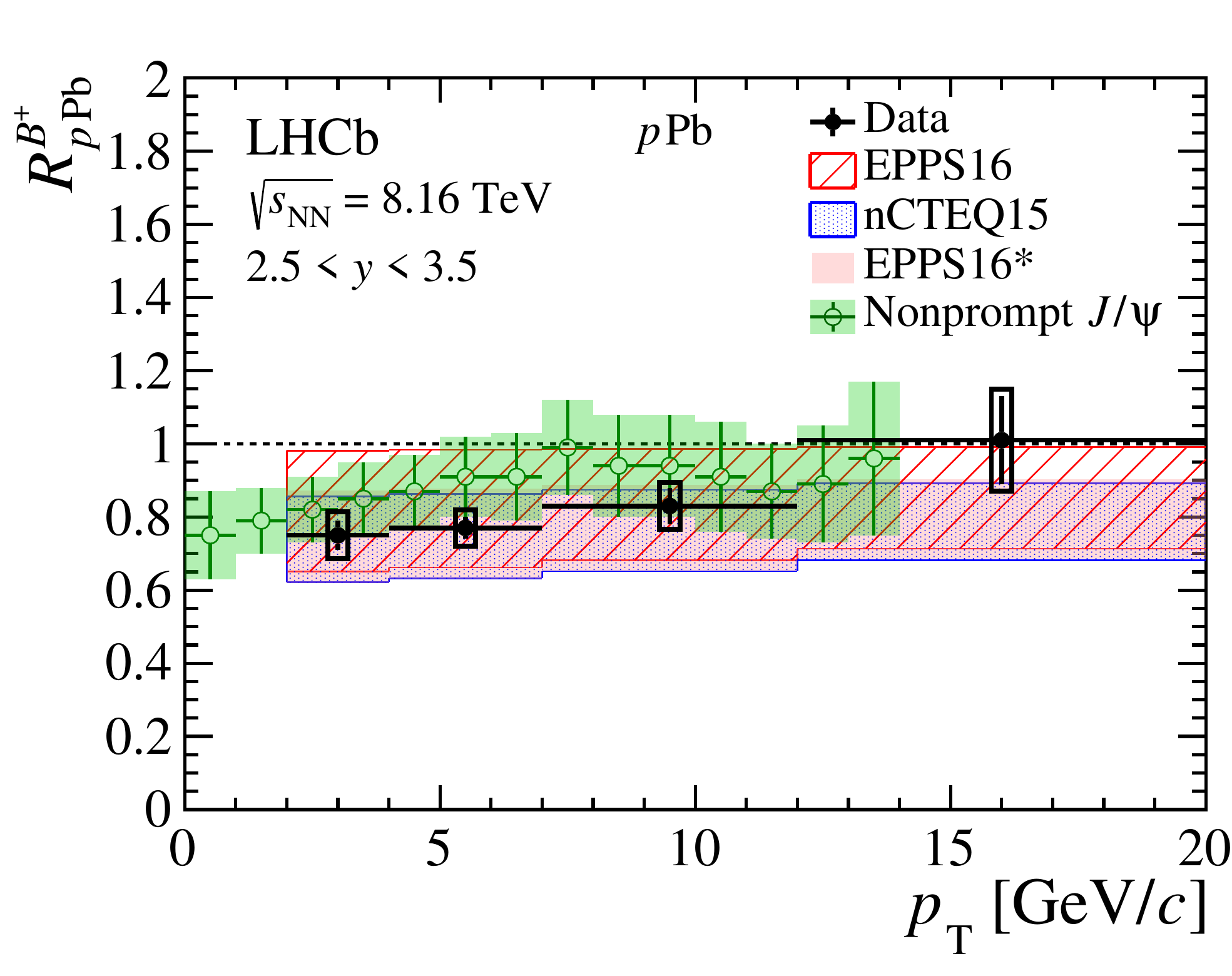}
  \includegraphics[width=0.49\textwidth]{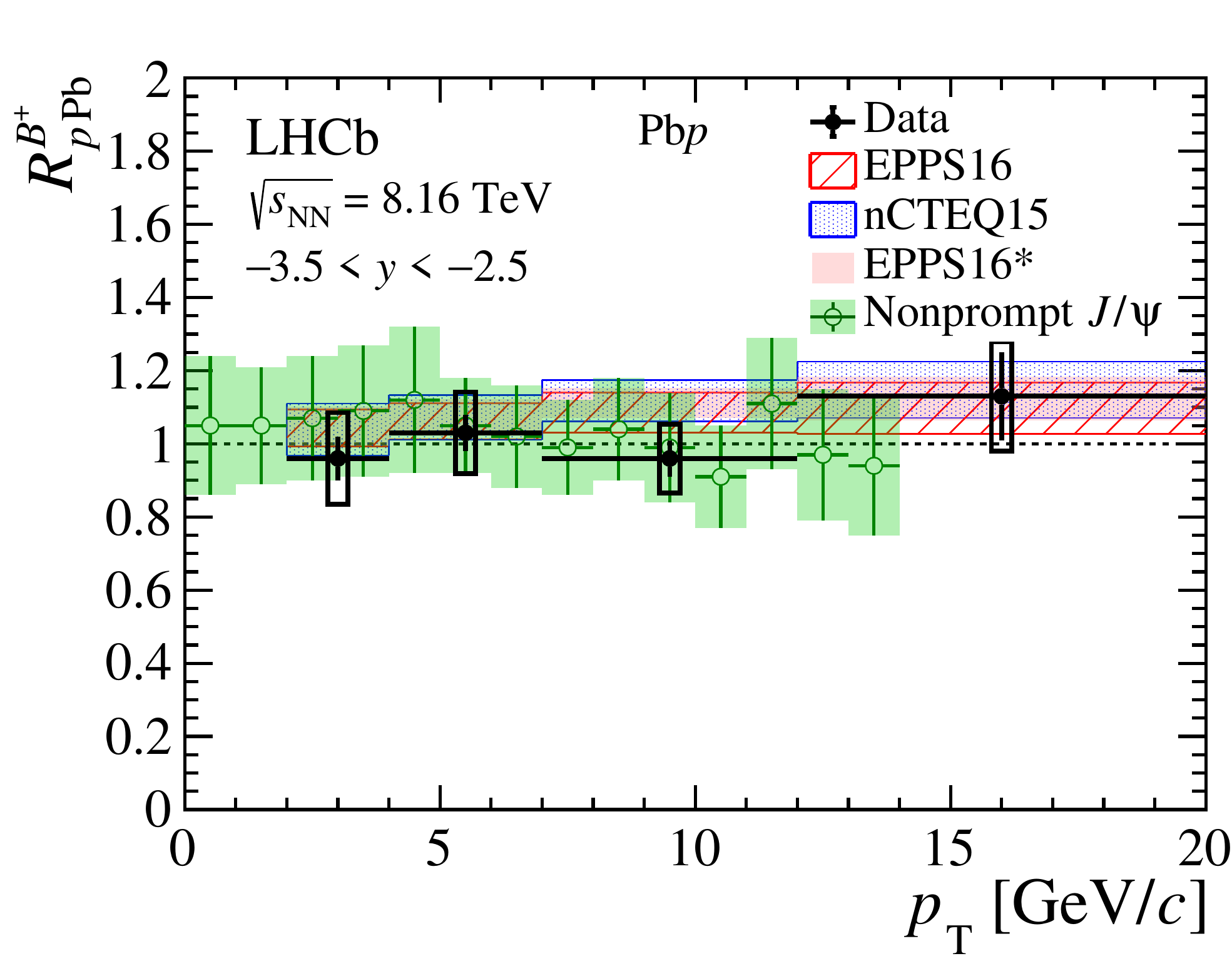}
  \caption{Nuclear modification factor, $R_{\pPb}$, for $\Bp$ mesons as function of (top) $y$ and as a function of \pt  in (bottom left) $\pPb$ and 
  (bottom right) $\Pbp$ compared with HELAC-onia calculations using different nPDF sets as well as with the measurement of
  $R_{\pPb}$ for nonprompt \jpsi production.
    For the data points, the vertical bars (boxes) represent the statistical (total) uncertainties.
    }
    \label{fig:RpA_Bplus}
\end{figure}

The $R_{p\,{\rm Pb}}$ modification factor for $\Bp$ production is shown in Fig.~\ref{fig:RpA_Bplus}, with the numerical values given in Table~\ref{tab:rpPb}. The
values are reported integrated over the considered \pt range for the two $y$ intervals, $-3.5<\ystar<-2.5$ and $2.5<\ystar<3.5$.
They are also given as a function of \pt for both $\pPb$ and $\Pbp$ collisions. 
For the $pp$ reference cross-section, an interpolation between existing $pp$ cross-section 
measurements by the LHCb collaboration at
7\tev~\cite{LHCb-PAPER-2013-004} and 13\tev~\cite{LHCb-PAPER-2017-037} is performed. A power-law function
is used following the approach of Refs.~\cite{LHCb-PAPER-2013-052,LHCb-PAPER-2015-058,LHCb-PAPER-2017-014}, 
which yields a prediction of $\Bp$ production at $\sqs=8.16\tev$ consistent with an extrapolation of the measured value at 
$\sqs=7\tev$ using a FONLL calculation~\cite{FONLL,Cacciari:2012ny}. The interpolation takes into account the
correlations provided in Ref.~\cite{LHCb-PAPER-2017-037}. The measurement of $R_{p{\,\rm Pb}}$ for nonprompt \jpsi production at the same collision
energy by the LHCb collaboration~\cite{LHCb-PAPER-2017-014} is also shown. 

At positive rapidity, a significant suppression by more than 20\% is observed integrating over the whole $\pt$ range,
whereas at negative rapidity, the result is consistent with unity. The measurement is also consistent with that of
nonprompt \jpsi production obtained in a similar kinematic range. The \pt-differential result at positive rapidity shows a
significant suppression, at the level of  25\% for the lowest \pt bin. The ratio tends to increase for high \pt, 
however, the current experimental uncertainties that grow also with \pt do not allow
to establish a significant $\pt$ dependence. At negative rapidity, all values are consistent with a nuclear
modification factor of one. The experimental data points are in good agreement with the three considered nPDF sets. At
positive rapidity, the experimental uncertainties are smaller than the nPDF ones for the integrated values as well as for
the three lowest \pt bins, whereas the experimental uncertainties are typically larger at negative rapidity. Under the
assumption that the dominance of nuclear modification is via nPDFs,  the results in the \pPb  sample provide constraints
that can be used in future nPDF fits. 

\begin{table}[!t]
\caption{
    Nuclear modification factor, $R_{\pPb}$, of \Bp production in $\pPb$ and $\Pbp$ collisions, in bins of \pt for the range $2.5<|y|<3.5$. The first uncertainty is
        statistical  and the second systematic.
   }
\centering
\scalebox{1.}{
\begin{tabular}{@{}lr@{$\,\pm\,$}r@{$\,\pm\,$}rr@{$\,\pm\,$}r@{$\,\pm\,$}r@{}}
    \toprule
$\pt\,\,(\mathrm{GeV}\!/c)$  & \multicolumn{3}{c}{$\pPb$}  &\multicolumn{3}{c}{$\Pbp$} \\
\midrule
                     $(\phantom{1}2,\phantom{1}4)$ &          0.75 & 0.04 & 0.05 &          0.96 & 0.06 & 0.11 \\
                     $(\phantom{1}4,\phantom{1}7)$ &          0.77 & 0.03 & 0.04 &          1.03 & 0.05 & 0.10  \\
                    $(\phantom{1}7,12)$ &          0.83 & 0.05 & 0.04 &          0.96 & 0.05 & 0.08 \\
                   $(12,20)$ &          1.01 & 0.12 & 0.07 &          1.13 & 0.12 & 0.09 \\
\midrule
                    $(\phantom{1}2,20)$ &          0.78 & 0.02 & 0.05 &          1.00 & 0.03 & 0.10 \\
\bottomrule
\end{tabular}
}
\label{tab:rpPb}
\end{table}

\section{Conclusions}

The differential production cross-sections of $\Bp$, $\Bz$ mesons and $\Lb$ baryons in proton-lead collisions at $\sqsnn=8.16\tev$ are
measured in the range $2<\pt<20\gevc$ within the rapidity ranges $1.5<\ystar<3.5$ and $-4.5<\ystar<-2.5$. The cross-sections and the derived nuclear modification factors and 
forward-backward ratios of $b$-hadron production are measured for the first time with exclusive decay modes at transverse
momenta smaller than the mass of the hadrons. They represent the first measurement
of beauty-hadron production with different exclusive decay channels in nuclear collisions in that kinematic regime. 
The results with fully reconstructed beauty hadrons confirm the significant nuclear suppression of
beauty-hadron production at positive rapidity measured via nonprompt \jpsi mesons. The observed experimental uncertainties at positive
rapidity are smaller than those achieved in a weighting of nPDFs with heavy-flavor data. Therefore, this measurement
can serve as a valuable input for future fits of nPDF, assuming that modifications of nPDFs are the dominant
source of nuclear effects in proton-lead collisions at the LHC. Finally, the unique measurement of $\Lb$ production
constrains the fragmentation of the beauty quark in a nuclear environment. The baryon-to-meson cross-section ratio in proton-lead  collisions  is
found to be compatible with the equivalent ratio measured in $pp$ collisions, and more data will be needed to study whether nuclear
effects modify beauty baryon and meson production differently.  These findings are important steps towards a better
understanding of heavy-flavor production in nuclear collision systems and will serve as an input for the characterization
of the quark-gluon plasma with heavy-flavor observables.


\section*{Acknowledgements}
\noindent 
We would like to thank Huasheng Shao for providing the HELAC-Onia  
theoretical predictions.
We express our gratitude to our colleagues in the CERN
accelerator departments for the excellent performance of the LHC. We
thank the technical and administrative staff at the LHCb
institutes.
We acknowledge support from CERN and from the national agencies:
CAPES, CNPq, FAPERJ and FINEP (Brazil); 
MOST and NSFC (China); 
CNRS/IN2P3 (France); 
BMBF, DFG and MPG (Germany); 
INFN (Italy); 
NWO (Netherlands); 
MNiSW and NCN (Poland); 
MEN/IFA (Romania); 
MSHE (Russia); 
MinECo (Spain); 
SNSF and SER (Switzerland); 
NASU (Ukraine); 
STFC (United Kingdom); 
NSF (USA).
We acknowledge the computing resources that are provided by CERN, IN2P3
(France), KIT and DESY (Germany), INFN (Italy), SURF (Netherlands),
PIC (Spain), GridPP (United Kingdom), RRCKI and Yandex
LLC (Russia), CSCS (Switzerland), IFIN-HH (Romania), CBPF (Brazil),
PL-GRID (Poland) and OSC (USA).
We are indebted to the communities behind the multiple open-source
software packages on which we depend.
Individual groups or members have received support from
AvH Foundation (Germany);
EPLANET, Marie Sk\l{}odowska-Curie Actions and ERC (European Union);
ANR, Labex P2IO and OCEVU, and R\'{e}gion Auvergne-Rh\^{o}ne-Alpes (France);
Key Research Program of Frontier Sciences of CAS, CAS PIFI, and the Thousand Talents Program (China);
RFBR, RSF and Yandex LLC (Russia);
GVA, XuntaGal and GENCAT (Spain);
the Royal Society
and the Leverhulme Trust (United Kingdom);
Laboratory Directed Research and Development program of LANL (USA).


\addcontentsline{toc}{section}{References}
\setboolean{inbibliography}{true}
\bibliographystyle{LHCb}
\bibliography{main,standard,LHCb-PAPER,LHCb-CONF,LHCb-DP,LHCb-TDR}

\newpage
\centerline
{\large\bf LHCb Collaboration}
\begin
{flushleft}
\small
R.~Aaij$^{29}$,
C.~Abell{\'a}n~Beteta$^{46}$,
B.~Adeva$^{43}$,
M.~Adinolfi$^{50}$,
C.A.~Aidala$^{77}$,
Z.~Ajaltouni$^{7}$,
S.~Akar$^{61}$,
P.~Albicocco$^{20}$,
J.~Albrecht$^{12}$,
F.~Alessio$^{44}$
% ...

{\footnotesize \it

% ...
$ ^{7}$Universit{\'e} Clermont Auvergne, CNRS/IN2P3, LPC, Clermont-Ferrand, France\\
% ...
$ ^{12}$Fakult{\"a}t Physik, Technische Universit{\"a}t Dortmund, Dortmund, Germany\\
% ...
$ ^{20}$INFN Laboratori Nazionali di Frascati, Frascati, Italy\\
% ...
$ ^{29}$Nikhef National Institute for Subatomic Physics, Amsterdam, Netherlands\\
% ...
$ ^{43}$Instituto Galego de F{\'\i}sica de Altas Enerx{\'\i}as (IGFAE), Universidade de Santiago de Compostela, Santiago de Compostela, Spain\\
$ ^{44}$European Organization for Nuclear Research (CERN), Geneva, Switzerland\\
% ...
$ ^{46}$Physik-Institut, Universit{\"a}t Z{\"u}rich, Z{\"u}rich, Switzerland\\
% ...
$ ^{50}$H.H. Wills Physics Laboratory, University of Bristol, Bristol, United Kingdom\\
% ...
$ ^{61}$University of Cincinnati, Cincinnati, OH, United States\\
% ...
$ ^{63}$Syracuse University, Syracuse, NY, United States\\
% ...
$ ^{77}$University of Michigan, Ann Arbor, United States, associated to $^{63}$\\
% ...
}
\end{flushleft}

\end{document}